\documentclass{aa}  

\bibpunct{(}{)}{;}{a}{}{,}

\usepackage{graphicx}
\usepackage{txfonts}
\usepackage{hyperref}
\hypersetup{
    colorlinks=true,
    linkcolor=blue,
    filecolor=magenta,      
    urlcolor=red,
    citecolor=blue
}

\usepackage{gensymb}
\usepackage{mathtools}
\usepackage{multicol}
\usepackage{float}
\usepackage{amsmath}
\usepackage{comment}
\usepackage[normalem]{ulem}

\RequirePackage{soul}
\RequirePackage{xcolor}

\begin{document}

    \title{Flare echos from relaxation
    shocks in perturbed
    relativistic jets}

    \titlerunning{Flare echos from relaxation
    shocks in perturbed
    relativistic jets}

    \author{G. Fichet de Clairfontaine,
          \inst{1}
          Z. Meliani
          \inst{1}
          and A. Zech
          \inst{1}
          }
          
    \authorrunning{G. Fichet de Clairfontaine, Z. Meliani and A. Zech}

    \institute{Laboratoire Univers et Théories, Observatoire de Paris, Université PSL, Université de Paris, CNRS, F-92190 Meudon, France \\
    \email{gaetan.fichetdeclairfontaine@obspm.fr} 
    }

   \date{Received January 14, 2022; accepted March 4, 2022}

  \abstract
   {One of the main scenarios to account for the multi-wavelength flux variability observed in relativistic jets of active galactic nuclei (AGNs) is based on diffusive shock acceleration of a population of relativistic electrons on internal shocks of various origins. Any complete AGN emission scenario has to be able to explain the wide range of observed variability time scales that change over several orders of magnitude between the radio and gamma-ray band. In addition to observations of flux variability, constraints are also provided by very-long-baseline interferometry (VLBI), which shows a large variety of moving and standing emission zones with distinct behavior.}
   {Combining dynamic hydrodynamic jet simulations with radiative transfer, we aim to characterize the evolution of stationary and moving emission zones in the jet and to study their multi-wavelength signatures through emission maps and light curves. We focus our study on flare events that occur during strong interactions between moving ejecta and stationary recollimation shocks. Such events are shown to lead to a significant perturbation of the stationary jet structure.}
   {We simulate relativistic jets with the magneto-hydrodynamic code \texttt{MPI-AMRVAC} and inject non-thermal particle distributions of electrons in shock regions. We follow the propagation of a moving shock and its interactions with a structure of standing re-collimation shocks in the jet. Synchrotron emission and radiative transfer are calculated in the post-processing code \texttt{RIPTIDE} for given observation angles and frequencies, assuming a turbulent magnetic field and taking into account light crossing effect.}
   {In the case of strong shock-shock interactions, we demonstrate the appearance of trailing components behind the leading moving shock. The latter destabilizes the jet, causing the emergence of oscillating standing shocks and relaxation shocks. Emissions from these regions can dominate the overall flux or lead to ``flare echos'' in the light curve. Another observational marker for the presence of relaxation shocks appears in time-distance plots of bright VLBI components of the jet. Our scenario provides a plausible explanation for radio VLBI observations of the radio-galaxy 3C\,111 where trailing components have been observed during a radio outburst event in 1997, and may be applicable to other sources with similar features.}
   {}

    
   \maketitle
%
%________________________________________________________________

\section{Introduction}
\label{sec: introduction}

% Introduction

%%%%%%%%%%%%%%%%%%%%
% Observation of relativistic  jet and variablities
%%%%%%%%%%%%%%%%%%%%
Relativistic jets in active galactic nuclei (AGNs) are among the most powerful phenomena in the universe. Their emission is observed across the entire electromagnetic spectrum, from the radio to the $\gamma$-ray band, and can exhibit extreme variability, especially at
high energies. The mechanisms behind this variability are still a matter of debate, since the fastest flaring events occur at spatial scales that remain inaccessible even to interferometric observations. The comparison of multi-wavelength (MWL) data on flares and flux variability with radiative models and jet models helps to narrow down the physical scenarios that may be at play.
Today, different scenarios are proposed that explain the observed variations in terms of internal or external shocks \citep{Marscher_1985ApJ...298..114M, Bottcher_2019, Lemoine_2019}, magneto-hydrodynamic instabilities \citep{Tammi_2009, Tramacere_2011} and magnetic reconnection \citep{Blandford_2017, Shukla_2020}, pulsar-like emission events close to the central black hole \citep{Aleksic_2014, Aharonian_2017}, or geometric effects, such as jet precession \citep{Raiteri_2017, Britzen_2019}. While several of these mechanisms may account for flaring events in different sources, the scenarios evoking internal shocks can provide a straightforward connection between time-dependent very-long-baseline interferometry (VLBI) data, which probe the structure of nearby relativistic jets, and MWL flares. 

%%%%%%%%%%%%%%%%%%%%%%%%%%%%%%%%%%%%%%%%
% Link variabilities with standing and moving shocks
%%%%%%%%%%%%%%%%%%%%%%%%%%%%%%%%%%%%%%%%
%%%%%%%%%%%%%%%%%%%%
% The standing chocs theory/Observations
%%%%%%%%%%%%%%%%%%%%
VLBI radio observations have revealed both moving and quasi-stationary features in different classes of AGNs \citep{Lister_2018, Lister_2021}. These bright ``radio knots'' have been seen from the radio to the millimeter band \citep{Lister_2009, Perlman_1999, Britzen_2010, Fromm_2011, Fromm_2013a, Fromm_2013b, Jorsatd_2013, Hervet_2016}, with counterparts up to the X-ray band in the most nearby sources \citep{Marshall_2002, Wilson_2002}. In the $\gamma$-ray range, these knots cannot be resolved, but the high-energy emission is usually attributed, at least in part, to compact regions inside the jet, due to the observed very rapid variability. Quasi-stationary knots may be interpreted as re-collimation shocks, naturally occurring in over-pressured jets \citep{Marscher_2008, Fromm_2013a, Hervet_2017, Fichet_2021}. This mechanism is for example widely accepted to explain the HST-1 radio knot in the nearby radio-galaxy M\,87 \citep{Stawarz_2006}. 

%%%%%%%%%%%%%%%%%%%%
% Moving shock
%%%%%%%%%%%%%%%%%%%%
Moving knots are observed in many sources \citep{Lister_2013, Walker_2018}, injected at the base of the jet or as a result of the ``detachment'' of previously
stationary knots, and can have superluminal apparent speed at large distances from the core. In some cases, new moving knots seem to be created by interactions within the jet. In their work, \cite{Kadler_2008} observed, in VLBI data of the radio galaxy 3C\,111, the displacement of ``trailing components'' that can emerge in the wake of bright, rapidly moving components. 

Interactions between moving knots and standing knots have been proposed to give rise to MWL flares \citep{Agudo_2012, Wehrle_2016, Kim_2020}, which can be interpreted as shock-shock interactions. 
The typical time scale and temporal shape of flares are wavelength-dependent. In the radio band, using a sample of 24 radio loud AGNs (BL Lacertae objects),  \citep{Nieppola_2009} shows that the variability time scale in the radio domain is close to 2.5 years. 
This variability time scale can decrease to hours in the X-ray band, due to much faster electron synchrotron cooling \citep{Massaro_2004}. 

While certain quasi-stationary VLBI knots in AGN jets are well understood, most stationary or moving features are still difficult to apprehend with the current models. \cite{Jorstad_2005} study the dynamics of 73 superluminal knots for various types of sources. They find a large variety of behaviour from purely ballistic motion, accelerated ones and evidence of trailing components which appear in the wake of rapidly moving knots. Such observations emphasize the need for a coherent scenario to interpret VLBI observations and to try to link them to the observed MWL variability. 

%%%%%%%%%%%%%%%%%
% Our model
%%%%%%%%%%%%%%%%%
In a previous study~\citep{Fichet_2021}, it was shown how flares can emerge in the radio band from shock-shock interactions for different jet configurations.
It was also seen that interactions between moving and standing shocks can have a non-negligible effect on the jet structure, leading to a temporary displacement of standing shocks. To further explore this effect, we concentrate here on the study of MWL flares resulting from strong interactions between moving shock waves and standing shocks, which will be shown to trigger secondary, trailing shocks with characteristic signatures. With this aim, we perform special relativistic hydrodynamic simulations of propagating shock waves in over-pressured jets subject to re-collimation shocks \citep[cf.][]{Fromm_2016}. The initial model presented by~\cite{Fichet_2021} has been extended to cover a larger wavelength range, up to the X-ray domain. 

 %%%%%%%%%%%%%%%%%
% Particles physics (injection/cooling)
%%%%%%%%%%%%%%%%%
To correctly treat the impact of radiative cooling on the particle population and its emission at high energies, electrons are injected only in shock regions and propagated along the jet.  
The spatial and temporal evolution of this electron fluid is treated in the hydrodynamic simulation, taking into account fluid advection and radiative and adiabatic energy losses \citep[cf.][]{van_Eerten_2010}.
 %%%%%%%%%%%%%%%%%
% radiation transfer and synthetic image
%%%%%%%%%%%%%%%%%
The radiative transfer equation is solved in post-processing \citep{Fichet_2021}, accounting for synchrotron absorption along a given line of sight (l.o.s.). Various methods have been proposed in the literature to take into account the time delays between signals emitted from different regions inside the jet. For the  present study, we start from the method proposed by \cite{Chiaberge_1999} and extend it to be applicable to a general flux map of a fully simulated jet and for arbitrary viewing angles.
  
%%%%%%%%%%%%%%%%%
% What new in this paper
%%%%%%%%%%%%%%%%%
So far, most hydrodynamic studies of shock-shock interactions in jets have focused on interactions of weak, moving perturbations with standing or moving shocks \citep{Mimica_2009, Fromm_2016, Vaidya_2018, Winner_2019, Fichet_2021, Huber_2021}. In this paper, the study of the interaction between strong, moving shocks with successive re-collimation shocks allows us to investigate the perturbation of the standing shock structure and the emergence of relaxation shocks, which can arise after strong shock-shock interactions.
 
 %%%%%%%%%%%%%%%%%%%%%%%%
 % Paper organization
 %%%%%%%%%%%%%%%%%%%%%%%%
 
In Section \ref{sec: numerical}, we summarize the numerical method implemented in the \texttt{MPI-AMRVAC} code to solve the relativistic hydrodynamic equations and the temporal evolution of relativistic electrons. Then we present shortly the \texttt{RIPTIDE} code (Radiation and Integration Processes with Time Dependence) that calculates the synchrotron flux and assures the treatment of the light crossing effect (LCE). 
In Section \ref{sec: results MPI} and Section \ref{sec: results rad}, we present our results respectively from a hydrodynamic and radiative point of view.  
Two observable features linked to the occurrence of relaxation shocks are described in Section \ref{sec: discussion} and their presence in already existing
data sets, e.g.\ from the source 3C\,111, is discussed. \\
Throughout this paper we will use natural units where the speed of light $c = 1$.

%%%%%%%%%%%%%%%%%%%%
% END %%%%%%%%%%%%%%
%%%%%%%%%%%%%%%%%%%%

\section{Model Description}
\label{sec: numerical}

%%%%%%%%%%%%%%%%
% Numerical method in use
%%%%%%%%%%%%%%%%
The fluid simulation is performed with the special relativistic magneto-hydrodynamic finite volume code \texttt{MPI-AMRVAC} \citep{Meliani_2007, Keppens_2012} using the Harten-Lax-van Leer-Contact (HLLC) Riemann solver \citep{Mignone_2006} with a third order reconstruction method \texttt{cada3} \citep{cada_2009}. The output, in the form of two-dimensional maps of a large set of variables for a series of time-steps, is treated in post-processing with \texttt{RIPTIDE}, to compute synthetic emission maps and light curves.

%%%%%%%%%%%%%%%%%%%%%%%%%%%%%%
% Fluid equations
%%%%%%%%%%%%%%%%%%%%%%%%%%%%%%
\subsection{Governing fluid equations}
\label{subsec: fluid equations}

The special relativistic hydrodynamic evolution of a perfect fluid is governed by the conservation of mass, momentum and energy, 
\begin{align}\label{eq: fluid equations}
%%%%%%%%%%%%%%%%%
\partial_{t} \, D+\nabla \cdot \left(D\,\vec{v}\right) & =  0, \\
%%%%%%%%%%%%%%%%%
\partial_{t} \, \vec{S} + \nabla \cdot \left(\vec{S} \vec{v}+p\right) & =  \vec{0}, \\
%%%%%%%%%%%%%%%%%
\partial_{t} \, \tau + \nabla \cdot \left(\left(\tau+p\right)\,\vec{v} \right) & =  0, 
%%%%%%%%%%%%%%%%%
 \end{align}
where $D=\gamma\rho$ is the lab-frame density, $\gamma$ is the Lorentz factor, $\rho$ is the density in the comoving frame, $p$ is the thermal pressure, 
%$\vec{u}=\gamma \vec{v}$ the four-velocity, 
$\vec{S} = D\,h\,\gamma\,\vec{v}$ the momentum and $\tau = \gamma D\,h - p - D$  a function of the total energy. The enthalpy $h = 1+\epsilon + p /\rho$ with $\epsilon$ the internal energy. 
%%%%
% EOS
%%%%%
To close the system of equations, we use the Synge equation of state \citep{Mathews_1971, Meliani_2004}. 

%%%%%%%%%%%%%%%%%%%%%%%%%%%%%%
% Electron evolution
%%%%%%%%%%%%%%%%%%%%%%%%%%%%%%
\subsection{Governing electron equations}
\label{subsec: e temporal evolution}
At shocks, a fraction $n_{\rm e} = \epsilon_{\rm e} \rho / m_{\rm p}$ of the electron population are injected in a power-law energy distribution (with $\epsilon_{\rm e}=0.01$ and $m_{\rm p}$ the proton mass) and an upper cut-off Lorentz factor $\gamma_{\rm e, max}$ that is arbitrarily set to an initial value of $10^7$. In the fluid simulation, this electron population is only characterized by $\gamma_{\rm e, max}$ and $n_{\rm e}$, while the explicit distribution over a power-law is carried out in post-processing.

The evolution of the electron number density is governed by the mass conservation equation,
\begin{equation}
    \label{eq: relativistic electrons  number density}
    \partial_{\rm t} \left(\gamma n_{\rm e}\right) +\nabla \cdot \left(\gamma n_{\rm e} \vec{v}\right)\,=\,0\,,
\end{equation}

To compute multi-wavelength light curves, we follow the temporal evolution of $\gamma_{\rm e, max}$ by resolving the evolution equation of the electron Lorentz factor $\gamma_{\rm e}$, taking into account adiabatic and synchrotron cooling,
%%%%%%
% EQUATION GAMMA_E GENERIC
%%%%%%
\begin{equation}
    \label{eq: relativistic electrons evolution}
    \dfrac{\textrm{d} \gamma_{\rm e}}{\textrm{d}t} = \dfrac{\gamma_{\rm e}}{3 n_{\rm e}} \dfrac{\textrm{d} n_{\rm e}}{\textrm{d} t} - \gamma_{\rm e}^2 \dfrac{\sigma_{\rm T} B^2}{6 \pi m_{\rm e}} \,,
\end{equation}
\noindent
where $\sigma_{\rm T}$ is the Thomson cross-section, $m_{\rm e}$ the electron mass, $n_{\rm e}$ the electron number density in the comoving frame, and $B$ the magnetic field strength. \\

%%%%%%
% Transition to MPI-AMRVAC
%%%%%%

For implementation in \texttt{MPI-AMRVAC}, we re-write Eq.~\ref{eq: relativistic electrons evolution} in the form of partial differential equations  \citep[cf.][]{van_Eerten_2010},
\begin{equation}\label{eq : relativistic electrons evolution MPI-AMRVAC}
    \partial_{\rm t}\left(\frac{\gamma \rho^{4/3}}{\gamma_{\rm e, max}}\right)+\nabla\cdot\left(\frac{  \gamma \rho^{4/3} }{ \gamma_{\rm e, max}}\vec{v}\right)=\frac{\sigma_{\rm T}}{6\,\pi\,m_{\rm e}\,c}\,\rho^{4/3}\,B^2\,.
\end{equation}
We set the minimum electron Lorentz factor to a fixed value of $\gamma_{\rm e, min} = 1$. For the treatment of the electron synchrotron cooling time step (Eq.~\ref{eq : relativistic electrons evolution MPI-AMRVAC}), we use a time steeping scheme \citep{Keppensetal_2020arXiv200403275K}. 
The values of $\epsilon_{\rm e}$ and $\epsilon_{\rm B}$ are set in such a way that the electron injection and the small scale magnetic field strength are in good
agreement with shock physics and do not influence the dynamics of the jet itself \citep[and references therein]{Zhang_2009, Vlasis_2011}.

\subsection{Shock detection}
\label{subsec: shock detection}
The electron distribution is reset according to the properties of the fluid at each shock. The accuracy of the shock detection method depends greatly on its ability to differentiate true shocks from compression waves. A robust method is required as shocks can be weak and therefore difficult to detect. In this work, we use the shock detection method described by \cite{Zanotti_2010} that is based on a close follow-up of the fluid Mach number. 

\subsection{Evolution of turbulent magnetic field strength}

In each shock region, a turbulent magnetic field is introduced that carries a fraction of the thermal energy $e_{\rm th}$, based on the following parametrization,
\begin{equation}
    B^2=4 \pi  \epsilon_{\rm B} \,e_{\rm th},
\end{equation}
where the free parameter $\epsilon_{\rm B}$ is set to $0.01$. \\

As in \cite{van_Eerten_2010}, the evolution of the advected small scale turbulent magnetic field strength is given by,
\begin{equation}
   \partial_{\rm t}\left(\frac{\gamma B^2}{\rho^{1/3}}\right)+\nabla\cdot \left(\frac{\gamma B^2 }{\rho^{1/3}}\vec{v}\right)\;=\;0\,.
\end{equation}

\subsection{Radiative transfer with \texttt{RIPTIDE} code}

The population of accelerated electrons that is injected and followed in the \texttt{MPI-AMRVAC} code, is distributed in the post-processing code \texttt{RIPTIDE} over a power-law distribution, as would be expected from shock acceleration,
\begin{equation}
    \dfrac{\textrm{d}n_{\rm e}}{\textrm{d}\gamma_{\rm e}} = K \gamma_{\rm e}^{-p} \,, 
    \label{eq: acc n1}
\end{equation}
\noindent
where $K$ is the normalization factor and $p = 2.2$ the typical power-law index for relativistic shocks. The above equation is valid for $\gamma_{\rm e, min} < \gamma < \gamma_{\rm e, max}$, with the latter evolving following Eq. \ref{eq: relativistic electrons evolution}.

Apart from the evolution of $\gamma_{\rm e, max}$ due to cooling, the description of radiation and radiative transfer follows the procedure detailed in~\cite{Fichet_2021}, which we summarize here for completeness sake.

The normalization factor $K$ is evaluated in each cell as follows,

\begin{equation}
    n_{\rm e} = K \int_{\gamma_{\rm e, min}}^{\gamma_{\rm e, max}} \gamma_{\rm e}^{-p} \text{d} \gamma_{\rm e} = \dfrac{K}{p - 1} \left(C_{\rm E}^{1 - p} - 1 \right) \cdot \gamma_{\rm e, min}^{1 - p} \,,
\end{equation}
where $C_{\rm E} = \gamma_{\rm e, max} / \gamma_{\rm e, min}$ and $n_{\rm e}$ are given by \texttt{MPI-AMRVAC}. \\

The radiative transfer equation is solved along a given line of sight (l.o.s) and with a given observation angle $\theta_{\rm obs}$, i.e.\ the angle between the observer l.o.s and the jet axis. Numerically, we obtain the synchrotron intensity for a given cell index \texttt{i} as,

\begin{equation}
    \label{eq: radiative transfer equation}
    I_{\nu; \,\rm \texttt{i}} = I_{\nu; \, \rm \texttt{i}-1} \exp{\left(-\tau_{\nu; \, \rm \texttt{i}}\right)} + S_{\nu; \, \rm \texttt{i}} \left(1 - \exp{\left(-\tau_{\nu; \, \rm \texttt{i}}\right)} \right),
\end{equation}
\noindent
where $S_{\nu} = j_{\nu}/\alpha_{\nu}$ the source function defined as the ratio of synchrotron emissivity $j_{\nu}$ and absorption coefficient $\alpha_{\nu}$, and $\tau_{\nu}$ the optical depth.\\ 

As the synchrotron intensity is emitted at the luminous distance $d_{\rm l}$ from the emitting surface $S_{\rm e}$, we consider the solid angle by calculating the synchrotron flux as measured from Earth,

\begin{equation}
    F_{\nu} = \left(\dfrac{S_{\rm e}}{d_{\rm l}^2}\right) \cdot \left( 1 + z \right) \cdot I_{\nu},
\end{equation}
\noindent
where $F_{\nu}$ is the flux in units of $\left[\textrm{erg} \cdot \textrm{cm}^{-2} \cdot \textrm{s}^{-1} \cdot \textrm{Hz}^{-1} \right]$. In this study, we assume a Hubble constant of $H_0 = 70~\text{km} \cdot \text{s}^{-1}  \cdot \text{Mpc}^{-1}$. For illustration
purposes, we choose a redshift equal to that of the radio galaxy 3C\,111 at $z = 0.049$ \citep{Truebenbach_2017}, since we will discuss VLBI observations of this source in Section~\ref{sec: discussion}.

\subsection{Light crossing effect}
\label{subsec: LCE method}

The light crossing effect (LCE) accounts for the different duration it takes photons emitted in different cells to propagate through the jet.

For a given choice of time steps with intervals $\Delta t'$ in the fluid simulation, the corresponding intervals in the observer frame are given as

\begin{equation}
    \Delta t_{\rm obs} = \dfrac{\Delta t'}{\delta \left(\theta_{\rm obs},~\gamma \right)} \cdot \left( 1 + z \right) \,,
    \label{eq: LCE equation}
\end{equation}

where $\delta \left(\theta_{\rm obs},~\gamma \right) = \left( \gamma \left( 1 - \beta \cos\left(\theta_{\rm obs} \right) \right)\right)^{-1}$ is the Doppler factor, determined for an average jet Lorentz factor $\gamma$. \\

To account for the LCE, we divide the jet in $N$ layers of equal width perpendicular to the observer's l.o.s.  The width of each layer corresponds to the distance traveled by the light during one time step in the observer's frame $c \, \Delta t_{\rm obs}$.
For a given source geometry and average jet Lorentz factor, the layers of simulated maps corresponding to different time steps are then matched to reproduce the radiative transfer with LCE. This method is presented in detail in Appendix \ref{sec: LCE appendix}. It should be noted that the full treatment of the LCE can require a large computation time as the number of emission layers increases with increasing Lorentz factor, decreasing viewing angle, and decreasing simulation time step. \\

The present model shows some similarities with the one presented by \cite{Fromm_2016}. In both cases the observed variability is explained through the shock - shock interaction scenario, where an initial perturbation is injected at the base of a jet that is structured with stationary recollimation shocks. However, in our scenario, we investigate a stronger shock-shock interaction, where the stationary recollimation shocks are strongly disturbed, to study the appearance of trailing features, which requires a high initial energy of the ejecta. Concerning the physic treated in our model, we extend the analysis towards a wider emission range, where the study by \cite{Fromm_2016} focuses on radio frequencies. Finally, the treatment of the particle is different in a way that we inject them directly on detected shocks. As we evaluate their temporal evolution though the jet simulations, this ensure the treatment of the radiative and adiabatic cooling.

\subsection{Setup of the variable jet scenario}
\label{subsec: scenario}

%%%%%%%%%%%%%%%%%%%%%%%%%%%
% Table for simulation parameters
%%%%%%%%%%%%%%%%%%%%%%%%%%%
\renewcommand{\arraystretch}{2}
\begin{table}
    \centering
    \begin{tabular}{c c c c}
        * & $\rho$ $\left[\text{cm}^{-3}\right]$ & $p$ $\left[\text{dyn} \cdot \text{cm}^{-2}\right]$ & $\gamma$ \\
        \hline
        ambient medium & $\rho_{\rm am} = 10^{3}$ & $p_{\rm am} = 1$ & $1$ \\
        \hline
        jet & $10$ & $1.5$ & $3$ \\
        \hline
        ejecta & $10$ & $1.5$ & $24$
    \end{tabular}
    \caption{Ambient medium, jet and ejecta parameters used in simulation.}
    \label{tab: sum up param}
\end{table}
\renewcommand{\arraystretch}{1}

\begin{figure*}[h!]
  \includegraphics[width = 2\columnwidth]{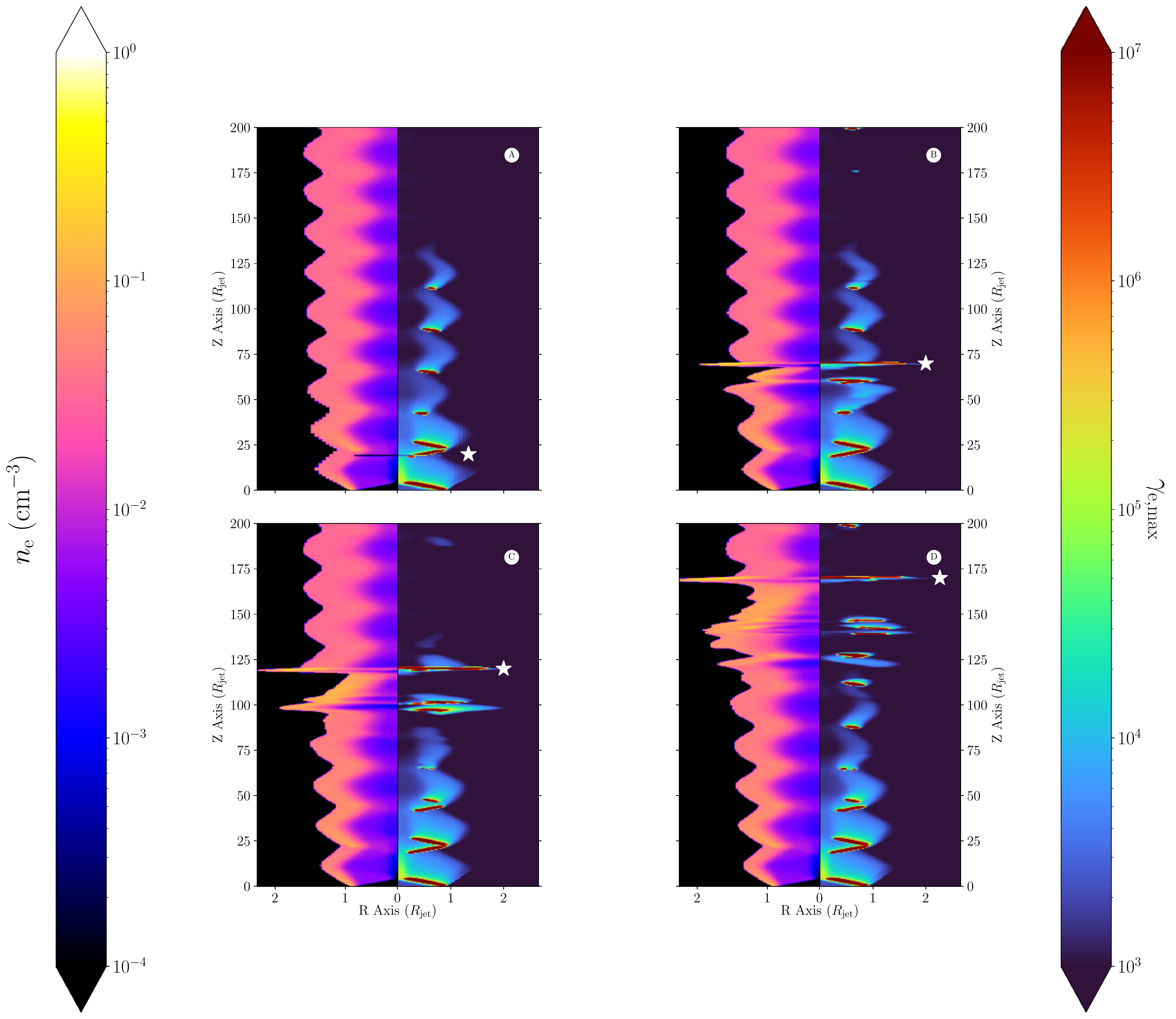}
  \caption{Snapshots of the jet with an injected perturbation. The jet is propagating from the bottom to the top along the Z-axis. The electron number density contour (in log-scale) is drawn on the left side and the maximum value of the electron Lorentz factor (also in log-scale) on the right side. Both the radial coordinate $R$ and $Z$ are given in $R_{\rm jet}$ units. The generated moving shock wave is located at $\sim 20~R_{\rm jet}$ (\textbf{A}), $\sim 70~R_{\rm jet}$ (\textbf{B}), $\sim 120~R_{\rm jet}$ (\textbf{C}) and $\sim 170~R_{\rm jet}$ (\textbf{D}) from the jet base and is marked with a white star. }
\label{fig: 1C MPI}
\end{figure*}

%%%%%%%%%%%%%%%%%%%%%%%%%%%
% context
%%%%%%%%%%%%%%%%%%%%%%%%%%%
To investigate the propagation of moving and relaxation shocks and their interactions with standing re-collimation shocks, we consider an over-pressured hydrodynamic jet surrounded by a uniform ambient medium.  Initially in a stationary state with standing shocks, the jet is destabilized by a component of the fluid at high Lorentz factor injected at its base (the ``ejecta''). The overall magnetization is supposed to be weak and only the magnetic field generated at shocks is being considered. This assumption is in accordance with the requirements for an efficient diffuse shock acceleration mechanism  on internal shocks \citep{Plotnikov_2018}. The magnetic field thus only affects electron cooling and not the jet dynamics. 

%%%%%%%%%%%%%%%%%%%%%%%%%%%
% Basic setting of over-pressured jet
%%%%%%%%%%%%%%%%%%%%%%%%%%%
For the jet geometry, we use values typical for radio loud relativistic AGNs, with a total kinetic luminosity of $L_{\rm kin} = 10^{46}~\text{erg} \cdot \text{s}^{-1}$ \citep{Ghisellini_2014} and a jet radius of $R_{\rm jet}\,=\,0.1\,{\rm pc}$ \citep{Biretta_2002}. We assume an initial Lorentz factor $\gamma_{\rm jet} = 3$. To trigger internal re-collimation shock patterns inside the jet, we assume that the jet is over-pressured compared to the ambient medium with a fixed pressure ratio of $p_{\rm jet} / p_{\rm am} = 1.5$ \citep{Gomez_1997} and a ratio of the rest mass density of $\rho_{\rm jet} / \rho_{\rm am} = 10^{-2}$. As we apply a constant density and pressure profile in the ambient medium and as the jet inlet is cylindrical, we expect to observe only an intrinsic jet opening angle of small amplitude \citep{Fichet_2021}. While this configuration does not reproduce the jet opening that is observed on the average \citep{Pushkarev_2017}, the objective here is to study the strong interaction of the moving shock with multiple stationary shocks. The cylindrical configuration permits to fulfill these conditions while maximizing the number of stationary shocks. The study of the behavior of moving shocks and shock-shock interactions in open jets is beyond the scope of this study and will be investigated in a dedicated application of the model to observational data.

%%%%%%%%%%%%%%%%%%%%%%%%%%%
% Moving shock wave
%%%%%%%%%%%%%%%%%%%%%%%%%%%

The perturbation injected at the jet base (with coordinates $\left( R ,\, Z \right) = \left( 0,\, 10 \right) ~ R_{\rm jet}$) is modeled as a cold spherical region with radius $R_{\rm ej} = R_{\rm jet} / 2$ and a density and pressure that are identical to those of the jet. However, its Lorentz factor is fixed to a larger value $\gamma_{\rm ej} = 24$. The ejecta will interact strongly with standing re-collimation shocks and disturb them. Its chosen characteristics allow the rapid formation of a detected moving shock in front of the ejecta. 
In this paper, since we are more concerned by the interaction between the moving shock and recollimation shocks, we consider a basic initial shape of the ejecta. Indeed, in our scenario, the shape for the ejecta doesn't change the characteristics of the resulting moving shock wave. The only important aspect is the initial energy flux of the ejecta. 
Its resulting luminosity is $L_{\rm ej} \simeq 10^{48}~\text{erg} \cdot \text{s}^{-1}$. 
All the parameters used in the simulations are summed up in Table~\ref{tab: sum up param}. \\

\section{Jet dynamics}
\label{sec: results MPI}

We first let the simulation of the over-pressured and supersonic jet evolve until it reaches a steady state with a pattern of standing shocks (Fig.~\ref{fig: 1C MPI}), similar to the one obtained in previous studies \citep{Fichet_2021}. A ``diamond'' like structure with a succession of compression and rarefaction regions arises within the jet as a result of the establishment of pressure equilibrium between the jet and the ambient medium (for more details see \citet{Wilson_1987, Marscher_2008, Fromm_2016, Hervet_2017, Fichet_2021}). When the steady state is reached, the distance between two successive re-collimation shocks is quasi constant and equal to $\delta Z_{\rm shock} = 2 \mathcal{M} \cdot R_{\rm jet} \simeq 22~R_{\rm jet}$, where the Mach number $\mathcal{M} = \gamma v / (\gamma_{\rm s} c_{\rm s})$ (with $c_{\rm s}$ the sound speed and $\gamma_{\rm s}$ its associated Lorentz factor). 

%%%%%%%%%%%%%%%%%%%%%%%%%%%
% detailed  description of shock wave pattern
%%%%%%%%%%%%%%%%%%%%%%%%%%%

Our detection method successfully identifies the standing shock regions, where relativistic electrons are injected. In these zones, the fluid has a bulk Lorentz factor of $\gamma = 3$ and can accelerate in rarefaction regions up to $\gamma = 4$. 

%%%%%%%%%%%%%%%%%%%%%%%%%
% Shock wave pattern and shock detection
%%%%%%%%%%%%%%%%%%%%%%%%%
Map \textbf{A} in Fig. \ref{fig: 1C MPI} shows the relativistic electron density and their maximum Lorentz factors of the jet shortly after injection of the perturbation. Since the strength of the shock waves decreases with distance from the jet inlet, the radial expansion of detected shock regions decreases with distance. Thus electron injection takes place mainly at the first few shocks and the electrons undergo radiative cooling as they move away from the shocks.

\begin{figure}
    \centering
    \includegraphics[width = \columnwidth]{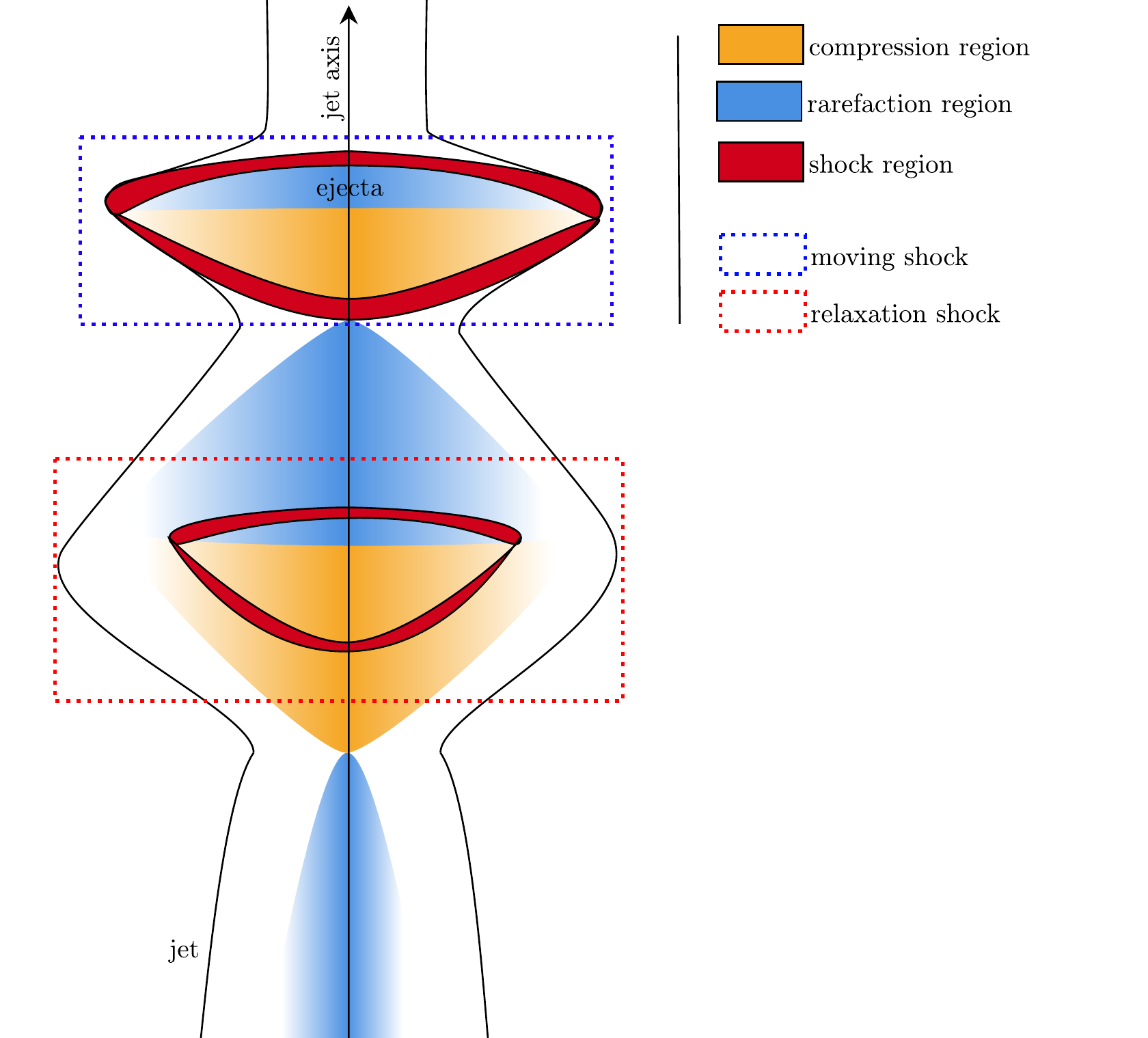}
    \caption{Schematic representation of the jet geometry during the propagation of the primary perturbation. We represent the formation of a relaxation shock in the trail of the moving shock. Scheme not to scale.}
    \label{fig: trail chain}
\end{figure}

%%%%%%%%%%%%%%%%%%%%%%%%%
% moving shock
%%%%%%%%%%%%%%%%%%%%%%%%%

%%%%%%%%%%%%%%%%%%%%%%%%%%%%%%%%%%%%%%%%
% Moving shock wave dynamic
%%%%%%%%%%%%%%%%%%%%%%%%%%%%%%%%%%%%%%%%

Once the jet has reached its steady state, a perturbation is injected at its base as described above. Fig. \ref{fig: 1C MPI}, \textbf{B} and \textbf{C} show the propagation of the fast ejecta, which interacts strongly with the jet material and triggers a moving shock. This main moving shock goes through a weak deceleration phase each time it crosses a standing re-collimation shock, which increases its thermal energy. A fraction of the jet material that is swept up and heated by the shock wave escapes radially toward the edge of the jet, which limits the amount of matter accreted by the moving shock and therefore its deceleration. As the moving shock runs into a rarefaction zone, it accelerates. This process is repeated each time the shock wave encounters a re-collimation shock and its following rarefaction zone. 

%%%%%%%%
% interation desciption with distance
%%%%%%%%
The first standing shock is only slightly disturbed by the passage of the moving shock. The second one is made to oscillate, leading to the emergence of rarefaction waves. In our setup, the third stationary shock is strongly destabilized. The interaction with the moving shock causes the emergence of a relaxation shock that propagates itself along the jet as a secondary moving, trailing shock. 

%%%%%%%%%%%%%%%%%%%
% Rarefaction zone and relaxation shock behind the shock wave
%%%%%%%%%%%%%%%%%%%%
As the primary moving shock sweeps up the jet material, it generates behind it a rarefaction zone (Fig.\ref{fig: 1C MPI}, \textbf{B}, \textbf{C}). Initially, the rarefaction zone expands along the jet axis between the moving shock and the standing re-collimation shock behind it. At the back of this zone, a relaxation shock emerges within the jet. During the first phase, until the moving shock reaches the third re-collimation shock, the relaxation shock remains a compression wave. Afterwards, this moving compression wave turns into a moving shock (Fig.\ref{fig: 1C MPI}, \textbf{B}, \textbf{C}). Behind this relaxation shock, the re-collimation shock undergoes a weak oscillation phase before it stabilizes. The strength of the relaxation shock increases with distance, disturbs more the following re-collimation shock and triggers another relaxation shock behind it. 
%%%%%%%%%%%%%%%
% More details
%%%%%%%%%%%%%%%
This succession of compression and rarefaction regions generated initially by the propagation of the perturbation is illustrated by the toy model in Fig. \ref{fig: trail chain}. At the end, when the moving shock and the trailing relaxation shocks leave the simulated jet, the latter recovers its quasi initial state (Fig.\ref{fig: 1C MPI}, \textbf{D}). 

%%%%%%%%%%%%%%%%%%%%
% END %%%%%%%%%%%%%%
%%%%%%%%%%%%%%%%%%%%

\section{Emission signature from moving and relaxation shocks}
\label{sec: results rad}
%%%%%%%%%%%%%%%%%%%%%%%%%%%%%%%%%
% General description
%%%%%%%%%%%%%%%%%%%%%%%%%%%%%%%%%
Applying the \texttt{RIPTIDE} post-processing code to the jet simulations, we calculated synthetic images and light curves for two observation angles ($\theta_{\rm obs} = 20 \degree$, $\theta_{\rm obs} = 90 \degree$ from the jet axis) and for four frequencies ($\nu \in \left[10^{10},\,10^{12},\,10^{14},\,10^{18}\right]~\text{Hz}$), with a time step of $R_{\rm jet} / c$. To investigate the impact of the LCE at $\theta_{\rm obs} = 90 \degree$, we present results on a selected time range with a smaller time step of $0.05~R_{\rm jet} / c$ in Appendix \ref{sec: LCE appendix}.

%%%%%%%%%%%%%%%%%%%%%%%%%%%%%%%%%%%%%%%%
% Synthetic image
%%%%%%%%%%%%%%%%%%%%%%%%%%%%%%%%%%%%%%%%
\subsection{Synthetic images}

%%%%%%%%%%%%%%%%%
% The synthetic image figure
%%%%%%%%%%%%%%%%%
\begin{figure}

    \centering

    \includegraphics[width = \columnwidth]{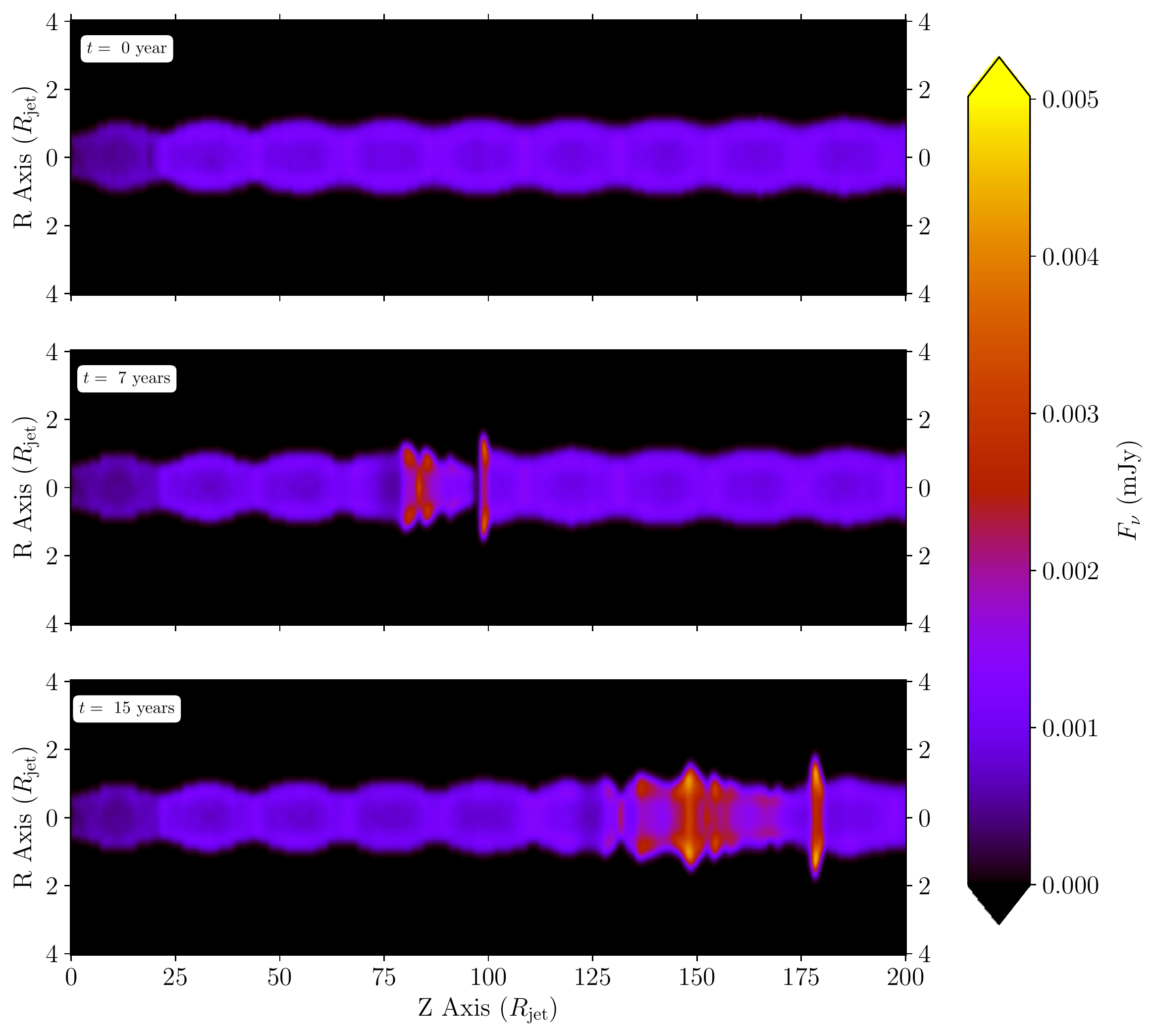}
    
    \includegraphics[width = \columnwidth]{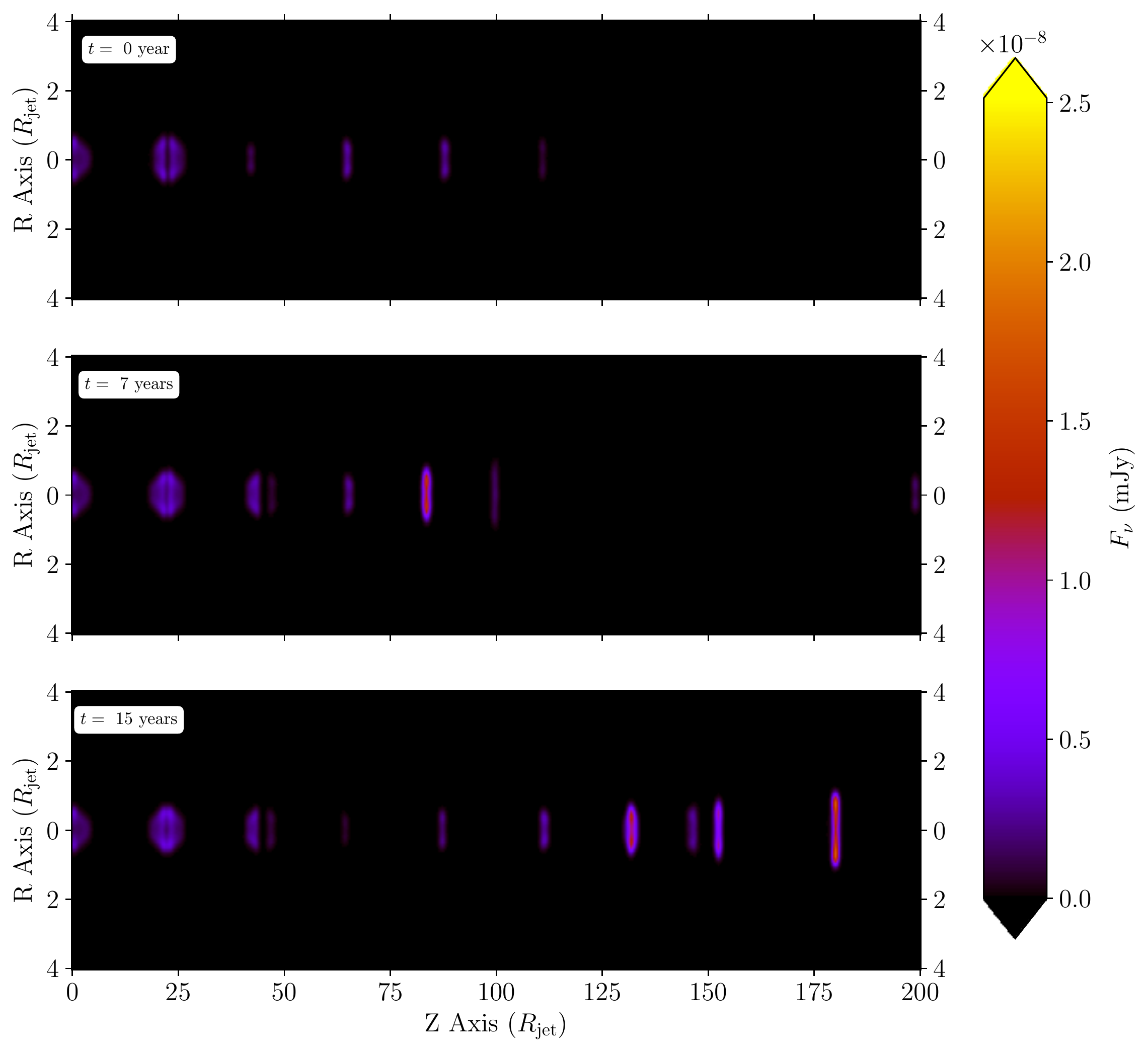}

    \caption{Snapshots : synchrotron emission maps of the jet before and after introduction of the ejecta, seen at two frequencies, $\nu = 10^{10}$ Hz (top) and $\nu = 10^{18}$ Hz (bottom). Each map represents the flux intensity in mJy units. The $R$ and $Z$-axis are given in $R_{\rm jet}$ units. These maps represent the flux in the co-moving frame for a viewing angle of $\theta_{\rm obs} = 90 \degree$ at three different co-moving times (see white boxes). The flux is evaluated at the distance of the radio galaxy 3C\,111}.
\label{fig: 1C sync map}

\end{figure}

%%%%%%%%%%%%%%%%%%%%%%%%%%%%%%%%%%
% stationary state
%%%%%%%%%%%%%%%%%%%%%%%%%%%%%%%%%%

As seen on Fig.~\ref{fig: 1C MPI}, which presents the jet in its initial, steady state, the shock strength decreases with distance from the jet inlet. Moreover, shocks are more pronounced towards the edge of the jet than along the jet axis, since the jet sound speed decreases outwards. Consequently, more electrons are injected near the jet edges and they remain advected along the steady shock pattern. They cool down by synchrotron emission as they propagate away from the shocks. Therefore, X-ray emission is most pronounced at shocks close to the jet inlet (Fig. \ref{fig: 1C sync map}, \textbf{bottom}), where the highly relativistic electrons with $\gamma_{\rm max} \simeq 10^{7}$ cool promptly. In the radio band, electrons continue to emit as they are propagating along the jet, given the longer cooling time for smaller Lorentz factors (Fig. \ref{fig: 1C sync map}, \textbf{top}).

%%%%%%%%%%%%%%%%%%%%%%%%%%%%%%%%%%
% moving shock (leading and relaxation)
%%%%%%%%%%%%%%%%%%%%%%%%%%%%%%%%%%

The emission from the moving shocks (leading and trailing) follows the same behaviour. Each time a moving shock interacts with a standing shock, the number of injected electrons grows, since $n_{\rm e} = 0.01~\rho / m_{\rm p}$. In the X-ray band, these electrons emit in a region close to the shock front and cool rapidly, while in the radio band, freshly injected electrons radiate during the propagation of the shock along the jet until they escape the jet at its edges. The escape time for the latter is close to $R_{\rm jet} / c \mathcal{M}$, the time taken by the moving shock to travel half the distance between two successive re-collimation shocks. \\
%%%%%%%%%%%%%%
%Relaxation shock 
%%%%%%%%%%%%%%%
In the synthetic maps, one also clearly sees the emergence of relaxation shocks after the third shock - shock interaction. If only one relaxation shock is visible at the beginning, their numbers increase at large distance. In the radio band, the emission surface associated to relaxation shocks increases and tends to dominate the emitted flux. In the X-ray band, individual relaxation shocks are clearly distinguishable.

%%%%%%%%%%%%%%%%%%%%%%%%%%%%%%%%%%%%%%%%
% Light curves
%%%%%%%%%%%%%%%%%%%%%%%%%%%%%%%%%%%%%%%%
\subsection{Light curves}\label{SubSec: Light curves}

\begin{figure*}[h!]
    \centering
    \includegraphics[width = 2\columnwidth]{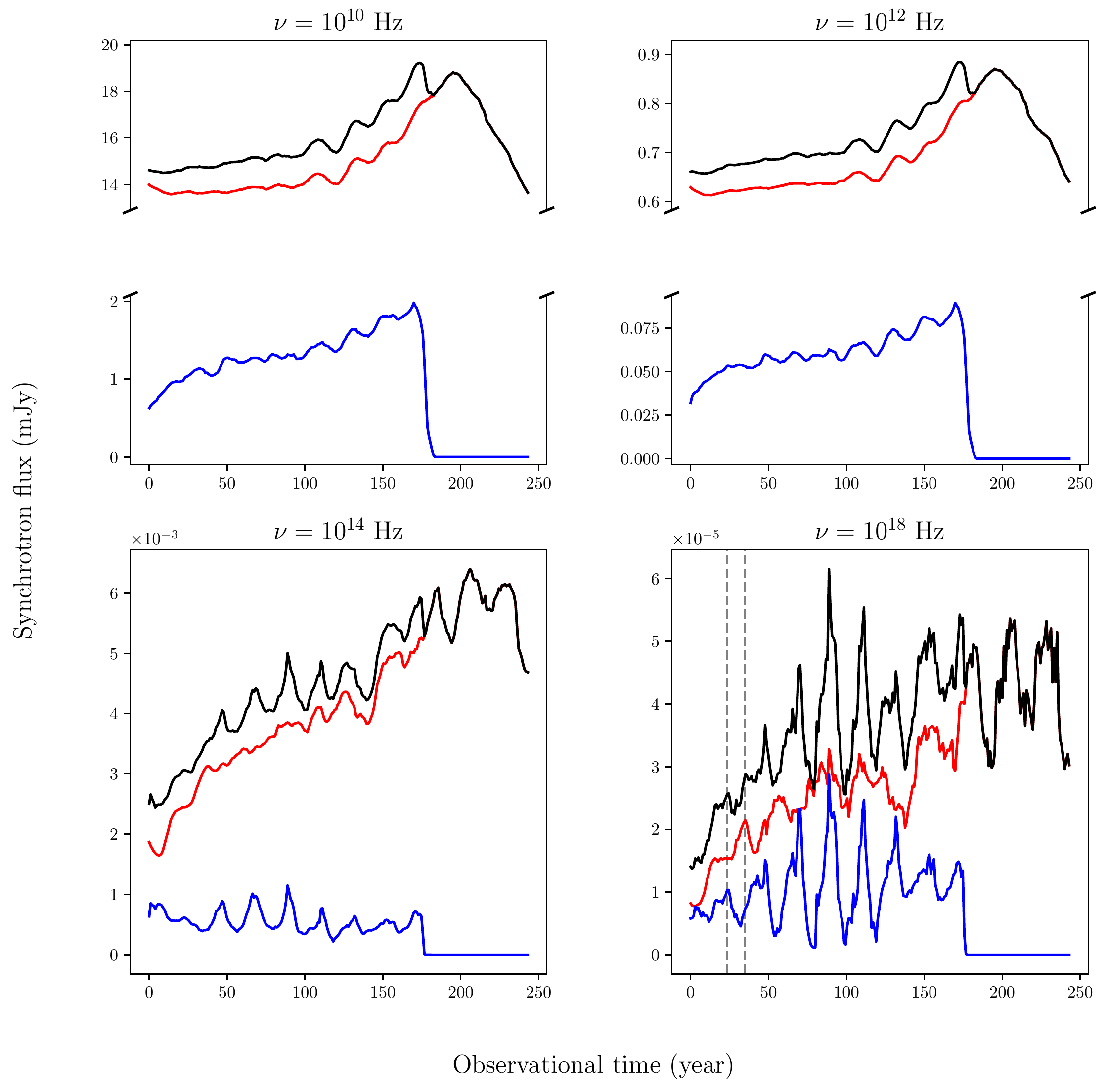}
    \caption{Light curves obtained by integrating the total synchrotron flux. The computation of the light curve is realized for four different frequencies (see titles). The overall flux is shown in black (upper curve), with the component from the original moving shock marked in blue (lowest curve) and the remaining jet emission in red. The flux is integrated from the injection time  of the ejecta (at $\text{t} = 0$ yr) until $t \sim 250 ~\text{yr}$. The dashed vertical lines in the plot with the X-ray light curve indicate the position of the first main flare (blue curve) and the associated flare echo from the perturbed jet (red curve). The flux is evaluated at the distance of the radio galaxy 3C\,111.}
    \label{fig: 1C - CL angle 90 - TD 0}
\end{figure*}

%%%%%%%%%%%%%%%%%%
%Overall description and behavior
%%%%%%%%%%%%%%%%%%

In Fig.~\ref{fig: 1C - CL angle 90 - TD 0}, we present synthetic multi-wavelength light curves computed for a viewing angle of $\theta_{\rm obs} = 90 \degree$ and for four different frequencies ($\nu \in \left[10^{10},\,10^{12},\,10^{14},\,10^{18}\right]~\text{Hz}$). Overall a continuous increase of the total observed flux (in black) is seen until the main moving shock and trailing relaxation shocks leave the simulated jet. Each time a moving shock crosses a re-collimation shock, a flare is triggered. This is particularly visible at high frequencies. As the LCE does not change the global shape of the light curve at $\theta_{\rm obs} = 90 \degree$ and with a time-step of $R_{\rm jet} / c$, this effect is not taken into account here. Its impact on the detailed structure of the flares is shown in Appendix. \ref{fig: LCE result}, where a higher temporal resolution was used. 

%%%%%%%%%%%%%%%%%%
% Low frequencies
%%%%%%%%%%%%%%%%%
At radio frequencies $\nu=\left[10^{10}, 10^{12}\right]~\text{Hz}$, the flux is dominated by the emission from electrons injected on re-collimation shocks (Fig.~\ref{fig: 1C - CL angle 90 - TD 0}, red line) and propagating along the jet.
Each time the leading moving shock interacts with a re-collimation shock, a new electron population starts contributing to the total flux, which thus increases continuously. After each interaction, depending on its strength, the disturbed re-collimation shock can radiate a remnant emission, which can lead to an elongation and asymmetry of the primary flare. The overall flux starts to decrease when the re-collimation shocks relax to their initial states and when relaxations shocks leave the observed jet fraction. \\

%%%%%%%%%%%%%%%%%%
% High frequencies
%%%%%%%%%%%%%%%%%

With increasing frequency, from the optical ($\nu = 10^{14}~\text{Hz}$) to the X-ray band ($\nu = 10^{18}~\text{Hz}$), the proportion of the flux coming from the moving shock increases to reach on the average $20~\%$ of the total flux in our setup (Fig. \ref{fig: 1C - CL angle 90 - TD 0}, blue line). In the light curves, rapid variability can be seen for each interaction between the moving and standing shocks. 
In the X-ray band, the typical variability time scale is given by the size of the stationary emission region, e.g.\ $5-10~R_{\rm jet} / c$ (in the co-moving frame) as at this frequency the synchrotron cooling time scale is shorter. It corresponds to the radial advection of the electrons by the shocked jet fluid towards the ambient medium along the moving shock front \citep{Meliani_2007b}.

%%%%%%%%%%%%%%
The proportion of the flux coming from standing shocks and relaxation shocks is still important at high frequencies. We observe, after each interaction, a radiative response of the jet, which can be classified in two types : during the two first shock-shock interactions, emission coming from perturbed standing shocks can be seen as delayed flares (or ``flare echos'', cf.~Fig.~\ref{fig: 1C - CL angle 90 - TD 0}, last panel, vertical lines). Starting from the third shock-shock interaction, delayed emission stemming from the moving relaxation shocks is added as well. \\

After the exit of the perturbation from the simulation box, relativistic electrons are still injected at moving relaxation shocks and provoke fast variability during interactions, before the flux returns to its quasi initial state (see Fig.~\ref{fig: 1C - CL angle 90 - TD 0}). 

\begin{figure}
    \centering
    \includegraphics[width = \columnwidth]{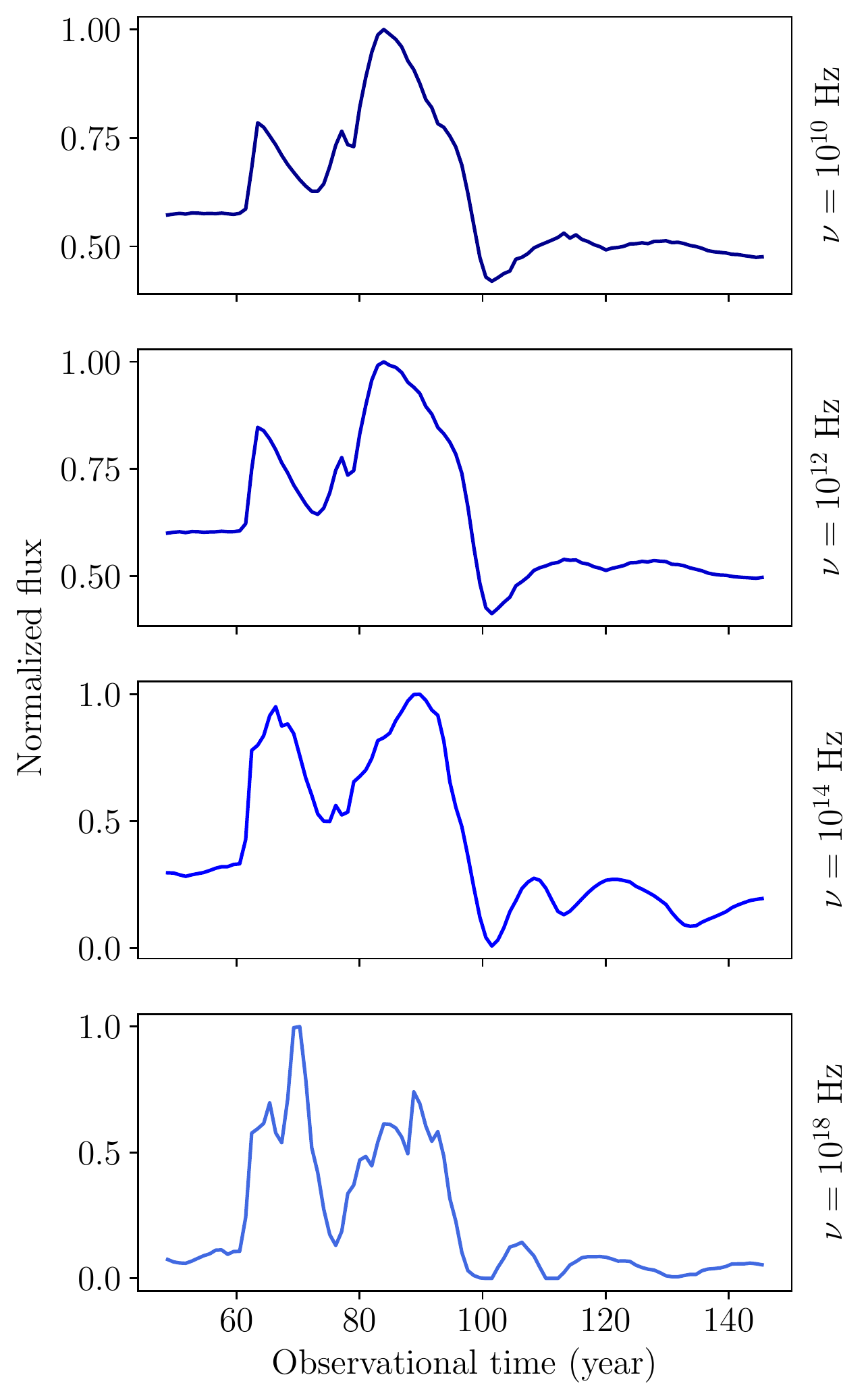}
    \caption{Light curves for four different frequencies integrated between $Z \in \left[75,~90\right]~R_{\rm jet}$, for an observational time between $t \in \left[4.8,~14.6\right]~\text{yr}$ and a jet viewing angle of $\theta_{\rm obs} = 90 \degree$. Fluxes are normalized to the maximum value.}
    \label{fig: zoom echo}
\end{figure}

In Fig.~\ref{fig: zoom echo}, we represent a zoom of the light curves centered around the third standing shock. 
We clearly distinguish the flare and its associated echo and avoid being dominated by the large
contribution from the emission of the surrounding jet.

At X-ray frequencies, due to the fast cooling, the flares from electron injection into the moving shocks show a sub-structure of several peaks (see Fig.~\ref{fig: zoom echo}, bottom). As the standing shocks are formed by two shock fronts corresponding to the compression and rarefaction region (cf. Fig.~\ref{fig: trail chain}), one observes two injection events for each passing of a moving shock. This structure is also visible in Fig.~\ref{fig: 1C - CL angle 90 - TD 0} (blue curve) in the X-ray light curve, for several flares coming from the principal moving shock. 
It should be noted that the width of the initial flare and flare echo in the radio band in Fig.~\ref{fig: zoom echo} is artificially reduced by the restricted size of the zoomed simulation box. As seen in the full simulation, flares are wider in the radio band.

%%%%%%%%%%%%%%%%%%%%
% END %%%%%%%%%%%%%%
%%%%%%%%%%%%%%%%%%%%

\section{Discussion}
\label{sec: discussion}

\subsection{Observable markers of jet relaxation.}
\label{subsec: fork VLBI}

% fork

\begin{figure*}
    \centering
    \includegraphics[width = 2\columnwidth]{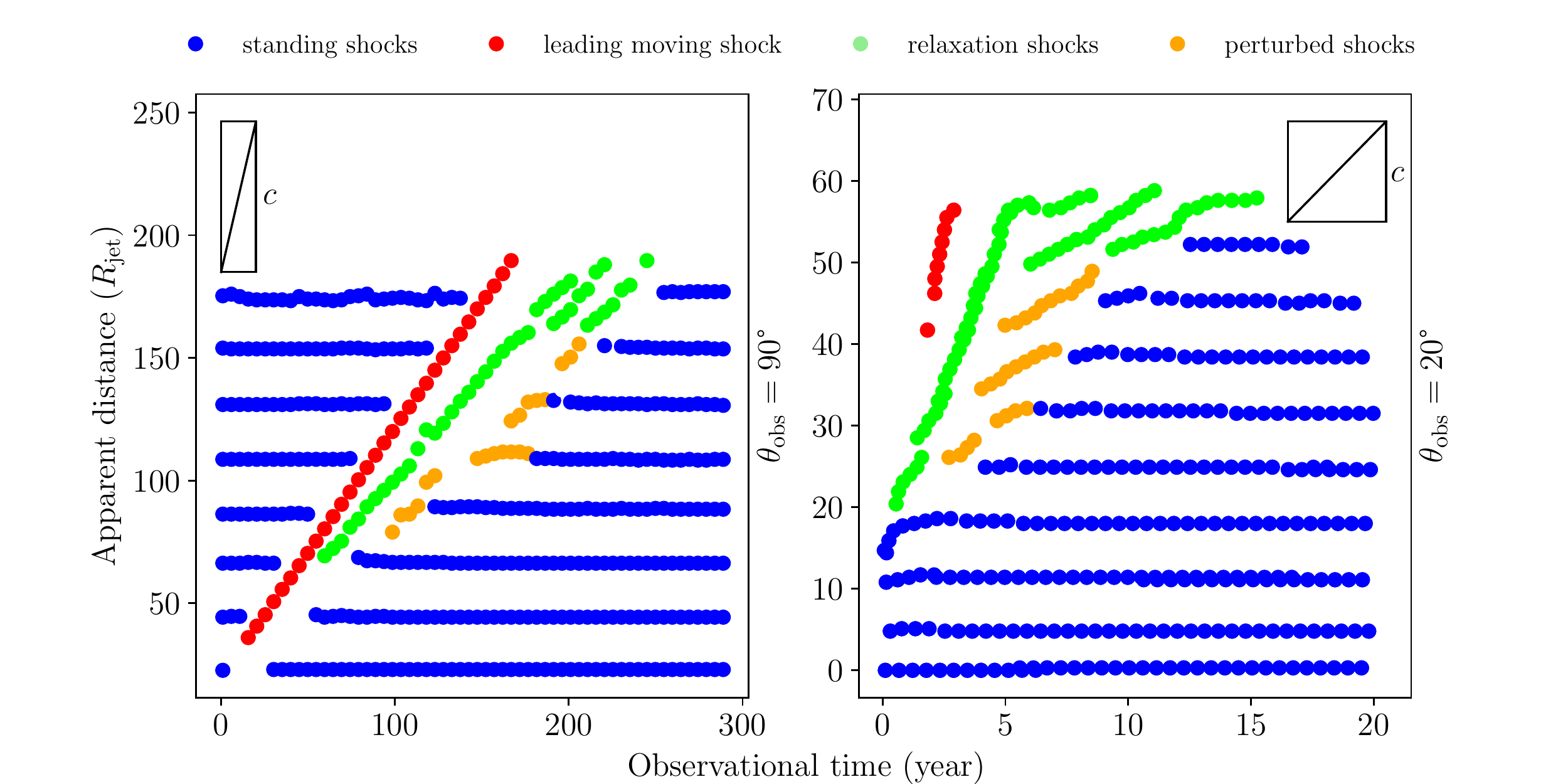}
    \caption{Apparent distance traveled by standing and moving knots in time. Each point represents, for a given time, the position of local maxima in the radio synchrotron flux at $\nu = 10^{10}$ Hz. For $\theta_{\rm obs} = 20 \degree$, the LCE is taken into account. The diagonal in the inset box corresponds to an apparent speed of $c$.}
    \label{fig: nodes distance vs time 10}
\end{figure*}

In our framework,we can classify the emission coming from the jet into four components. First the continuous emission from the steady jet which corresponds to the more or less extensive emission coming from electrons injected at standing shocks. Secondly, the emission from the moving shock leading the ejecta, which causes strong variability at high frequencies, due to fast-cooling electrons, as the ejecta propagates through the jet. Then, the remnant emission coming from oscillating standing shocks following an interaction with the ejecta. Finally, the emission from relaxation shocks trailing the leading shock. This emission can be extended over a long time. 

The temporal evolution of the jet internal shocks structure at $90$° and $20$° in the presence of fast-moving ejecta is shown in Fig. \ref{fig: nodes distance vs time 10}, where the apparent distances from the base of the jet of the brightest regions in the radio emission maps are indicated as a function of time in the observer frame. 
The regular structure of the equally spaced standing shocks corresponds to the blue points. The leading moving shock (red points) is identified by a line of constant slope, corresponding to the apparent speed of the ejecta. 

In our simulations, the two first interactions with standing shocks cause remnant emissions which are clearly visible (especially in the X-ray band) in the light curves (see Fig. \ref{fig: 1C - CL angle 90 - TD 0}). At the third interaction, the resulting oscillation is sufficiently strong to lead to the emergence of a relaxation shock with a lower speed corresponding to a second diagonal line with a smaller slope in Fig.~\ref{fig: nodes distance vs time 10} (green points). The velocity of the relaxation shock is always greater than the jet velocity; an upper limit on the jet velocity can thus be determined by analyzing the velocity of such relaxation shocks. 

As can be seen in Fig.~\ref{fig: nodes distance vs time 10} for a viewing angle of 90$^{\circ}$, the emergence of a relaxation shock at the intersection of a moving shock and a stationary shock is visible as a characteristic ``fork'' in the distance vs. time plot. At $20$°, this feature seems more difficult to detect (see right part of Fig. \ref{fig: nodes distance vs time 10}), due to higher opacity in the radio band and the limiting spatial resolution. 

% echo
Flare echoes may arise from the oscillation of perturbed standing shocks or from the propagation of relaxation shocks. This can be see in Fig. \ref{fig: 1C - CL angle 90 - TD 0} and more clearly in Fig. \ref{fig: zoom echo}, where a first flare from the shock-shock interaction occurs after 60 years in the observer frame, followed by a second flare due to a relaxation wave 20 years later, for this specific event. In our simulations, a relaxation shock appears after the third interaction. From this point on, it is difficult to distinguish, in the light curve, the remnant emission from oscillating standing shocks and emission from relaxation shocks. A non negligible emission counterpart due to either of the two mechanisms is expected at all frequencies and may be detected either as a separate flare echo or, if the initial flare has not ended as the secondary emission arises, in the form of a longer flare with a high asymmetry. The latter would be expected in particular in the radio band, where longer cooling times lead to a slower decay of the initial flare. \\

\subsection{Observations of trailing shocks in 3C\,111}
% fork and echo

Relaxation shocks may be identified with the trailing shocks that have already been detected in VLBI radio observations of several sources. Many of these trailing components do seem to appear after interactions of a moving 
radio knot with standing knots. Trailing components have apparent velocities that are always lower than that of the leading component.

%%%%%%%%%%%%%%%%%
% General description of 3C\,111
%%%%%%%%%%%%%%%%%%
A prominent example is the nearby Fanaroff-Riley II radio galaxy 3C\,111, at a redshift of $z = 0.049$ \citep{Truebenbach_2017}. Its parsec scale jet has been observed in the radio band \citep{Kadler_2008}, in X-rays \citep{Marscher_2006, Fedorova_2020} and up to the $\gamma$-ray band \citep{Grandi_2012}.
%%%%%%%%%%%
% Radio observations: shocks stationary and moving
%%%%%%%%%%%
Long term VLBI monitoring of 3C\,111 reveals the existence of bright superluminal knots \citep{Preuss_1988, Kadler_2008, Schulz_2020}. Recent observations by \citet{Schulz_2020}, and VLBA polarization monitoring \citep{Beuchert_2018}, suggest the presence of two stationary knots, \textbf{C2} and \textbf{C3}, located respectively at $\sim0.1~\text{mas}$ and $3~\text{mas}$ from the core.

%%%%%%%%%%%%%%
% Detection of the moving shock E and F 
%%%%%%%%%%%%%%
During the period of 1996 to 1997, radio monitoring with the Very Long Baseline Array (VLBA) 2cm Survey and the MOJAVE monitoring programs of detected two newly formed moving knots (named \textbf{E} and \textbf{F}), \citep{Kadler_2008}, emerging at the position where later on, during the 1999 observational campaign, the stationary knot \textbf{C2} was detected \citep{Beuchert_2018}. The appearance of knots \textbf{E} and \textbf{F} are associated with a bright radio flare event, which shows a strong asymmetry. 

%%%%%%%%%
% ZM I moved up (here now)
%%%%%%%%%
Additionally, \cite{Schulz_2020} (Fig.~1) have detected in 2007 a second radio flare event with a complex flare structure. The authors suggest that trailing components are present and find some evidence of similarities between this event and the one in 1997, although the characteristic fork signature cannot be seen in this case. The radio flare seems correlated to the traveling of moving knots near the first recollimation shock \textbf{C2}. 

%%%%%%%%%%
% Interpretation  E and F
%%%%%%%%%%%
\cite{Kadler_2008} interpreted the trailing component \textbf{F} as a reverse shock formed behind the leading knot \textbf{E} according to the scenario proposed by \citep{Perucho_2008} using 1D numerical hydrodynamic simulations.

%%%%%%%%%%%%%%
% Detection of the E->(E1,E2,E3)
%%%%%%%%%%%%%%
Further observations of the 1999 monitoring of the moving knot \textbf{E} show that at the position of the standing radio knot \textbf{C3}, the moving knot \textbf{E} splits up into a principal component \textbf{E1} and three trailing components (\textbf{E2}, \textbf{E3}, \textbf{E4}) \citep{Kadler_2008}.
%%%%%%%%%%%%%%%%
% Interpretation E1,E2,E3,E4
%%%%%%%%%%%%%%%%
The emergence of the these trailing features are interpreted by the authors following the scenario of the formation of conical moving shocks due to Kelvin-Helmholtz instabilities triggered solely by the propagation of an ejecta \textbf{E} within a hydrodynamic jet \citep{Agudo_2001}.
%%%%%%%%%%%%%%%%%%%%%%%%%%
During this phase no noticeable flares were detected in association with the knots (\textbf{E1}, \textbf{E2}, \textbf{E3}, \textbf{E4}). 

\subsection{Interpretation with the relaxation shock scenario}
%%%%%%%%%%%
% Scenario for E and F moving knot
%%%%%%%%%%%
In our proposed scenario, we interpret the knot \textbf{E} as a leading moving shock formed by an ejecta and \textbf{F} as the first standing shock perturbed by the leading moving shock (Fig.\ref{fig: 1C - CL angle 90 - TD 0}). After the interaction, the standing shock relaxes and retrieves its initial position. We can understand the linear flux increase in component \textbf{E} (cf. Fig. 6 in \cite{Kadler_2008}) as a consequence of continuous injection of electrons at the moving shock. The fast decrease of the emission from knot \textbf{F} might be better explained by a remnant emission coming from a perturbed standing shock, than by a relaxation shock. Such displacements of perturbed standing shocks can be seen in Fig.~\ref{fig: nodes distance vs time 10} (orange points) with an apparent speed close to the speed of light. It should be noted that, for small viewing angles, such perturbed shocks are difficult to distinguish from relaxation shocks, due to strong Doppler boosting, combined with the LCE.

%%%%%%%%%%%
% Scenario for E and C3 interaction
%%%%%%%%%%%
As the ejecta \textbf{E} is accelerated along the jet and interacts with \textbf{C3}, several trailing components, which we interpret as relaxation shocks, can be seen with a large variety of velocities. In our simulations, similar fork events can be seen at large distances from the core with an appearance of several relaxation shocks.
The large number of relaxation shocks might be explained by the strength of the interaction of the main moving knot \textbf{E}  with standing knot \textbf{C3}. The lower flux measured might be linked to a lower jet density in this region, as the moving shocks propagate in an expanding jet \citep{Beuchert_2018}. \\

The main difference between our scenario and the scenario proposed by \cite{Agudo_2001} is the presence of stationary shocks in the jets. The presence of stationary components allows the emergence of relaxation shocks during their oscillation after strong interaction. While their appearance is also possible in the absence of stationary components under certain conditions, interactions with the latter lead to a larger diversity of trailing components. Stationary components and their apparent drifting after interactions with a perturbation are observed in many sources with VLBI observations \citep{Kim_2020, Lico_2022}. The oscillation or drifting of stationary features can lead to additional emission counterparts and can help to explain observations as in the case of 3C\,111.

\subsection{Further examples for applications}
%%%%%%%%%%%%%
% 3C\,120 of shock-shock interaction
%%%%%%%%%%%%%
Another example, where our scenario may apply, is given by VLBI observations of the broad line radio galaxy 3C\,120. \citet{Gomez_2001} identified several trailing components triggered by an ejecta from the radio core in the 1998 VLBA data, which they ascribe to the result of dynamically induced recollimation shocks, again following \citet{Agudo_2001}. A later data set of VLBA monitoring data \citep{Jorstad_2017} then revealed three stationary knots at a distance of $<0.5$\,mas from the radio core, i.e.\ within the region where the trailing shocks arise. If these knots can be identified with stationary shocks that were already present, but undetected, in 1998, our scenario of relaxation shocks triggered by interactions between a primary moving shock and one or several standing shocks would apply.

%%%%%%%%%%%%%
% Other cases of fork signature
%%%%%%%%%%%%%
More generally, observations of trailing components are not limited to the prominent cases such 3C\,111 or 3C\,120. \cite{Jorstad_2005} have identified several sources with trailing components that may be signs of the presence of relaxation shocks. As an example the radio source PKS 0823+033 and 4C +10.45 show characteristic fork signature of relaxation shocks (Appendix \ref{sec: trailing candidates})  

%%%%%%%%%%%%%
% Relaxation flare
%%%%%%%%%%%%%
In addition to the study of VLBI maps, relaxation shocks should also be detectable as flare echos. As was seen in Fig.~\ref{fig: zoom echo}, such temporal signatures are more easily distinguishable at high energies, as in X-ray light curves. Compared to the radio band, flares are sharper at higher frequencies, since self-absorption is negligible and the cooling time scales shorter in comparison with the dynamical time. X-ray flare echoes or flare asymmetries might provide evidence for relaxation shocks or remnant emission from perturbed standing shocks in blazars, where the small observation angles make the observation of fork events in VLBI very difficult. However, the identification of such events might be complicated by the impact of the LCE on the light curve. 

%%%%%%%%%%%%%%%%%%%%
% END %%%%%%%%%%%%%%
%%%%%%%%%%%%%%%%%%%%

\section{Conclusion}
\label{sec: conclusion}
%%%%%%%%%%%%%%%%%%
% We are the best
%%%%%%%%%%%%%%%%%%
We have investigated, for the first time, the emergence of relaxation shocks resulting from the interaction of strong moving and standing recollimation shocks in relativistic jets.  

%%%%%%%%%%%%%%%%%%
% How: with SR-MHD simulation
%%%%%%%%%%%%%%%%%%
We carried out high resolution special-relativistic hydrodynamic simulations using the \texttt{MPI-AMRVAC} code of over-pressured jet subject to a strong variability at its inlet.
%%%%%%%%%%%%%%%%%%%
% RIPTIDE is the best in the market
%%%%%%%%%%%%%%%%%%%
We developed a new radiative transfer code \texttt{RIPTIDE} handling the light crossing effect and able to post-process large scale high resolution SR-HD simulations of AGN jets for different observation angles and to compute synthetic synchrotron images and light curves at various frequencies.\\
%%%%%%%%%%%%%%%%%%
% Our best results
%%%%%%%%%%%%%%%%%%
Based on these simulations, we propose a scenario able to reproduce the formation and the propagation of transient trailing shocks observed in VLBI observations. In our scenario, such features are induced by the strong interaction between shocks inside the jet. We showed that during such interactions, the standing shocks start to oscillate around their equilibrium position. For sufficiently strong oscillation events, new moving shocks are released that follow the leading one. 

The synthetic synchrotron light curves help us to distinguish between flares coming from interactions of the leading moving shock and additional flares associated with perturbed standing shocks and relaxation shocks. These additional components can lead to a high asymmetry of flares or flare echos in the observed light curve. 

A qualitative comparison of the expected observable features of the relaxation shock scenario with publicly available data of the radio galaxy 3C\,111 seems promising. We propose a new coherent scenario to explain the appearance of different trailing components in VLBI observations an emission counterpart in radio light curves. Trailing components are also seen in other radio galaxies and  that might again be explained with our scenario.

\begin{acknowledgements}
The authors thank the anonymous reviewer for his peer review. Computations of the SR-HD results were carried out on the OCCIGEN cluster at CINES \footnote{\url{https://www.cines.fr/}} in Montpellier (project named \texttt{lut6216}, allocation \texttt{A0090406842} and \texttt{A0100412483}). Radiative outputs given by the \texttt{RIPTIDE} code were carried out by the MesoPSL cluster at PSL University\footnote{\url{http://www.mesopsl.fr/}} in the Observatory of Paris. This work was granted access to the HPC resources of MesoPSL financed by the Region Ile de France and the project Equip@Meso (reference \texttt{ANR-10-EQPX-29-01}) of the programme Investissements d’Avenir supervised by the Agence Nationale pour la Recherche. This research has made use of data from the MOJAVE database that is maintained by the MOJAVE team \citep{Lister_2018, Lister_2021}. 
\end{acknowledgements}

%%%%%%%%%%%%%%%%%%%%
% END %%%%%%%%%%%%%%
%%%%%%%%%%%%%%%%%%%%

\bibliographystyle{aa.bst} 
\bibliography{biblio.bib}

\appendix

\section{Light crossing effect}
\label{sec: LCE appendix}

\subsection{Implementation and test on a toy model.}
\label{subsection: LCE toy}
The method we have implemented to account for the LCE is an extension of the procedure described by \cite{Chiaberge_1999}, which was limited to a simple geometry of the emission region, seen by a comoving observer under an angle of $90^\circ{}$.

The implementation of the LCE requires knowledge of the simulation time step $\Delta t'$ and average bulk Doppler factor of the jet. Depending on these parameters, we divide the simulation box, projected in the observer's frame for a given viewing angle, in $N$ layers perpendicularly to the observer's l.o.s., such that the width of each layer equals the distance light travels in one time step in the observer frame $\Delta t_{\rm obs}$ (cf. Eq.~\ref{eq: LCE equation}).

\begin{figure}[h!]
    \centering
    \includegraphics[width = \columnwidth]{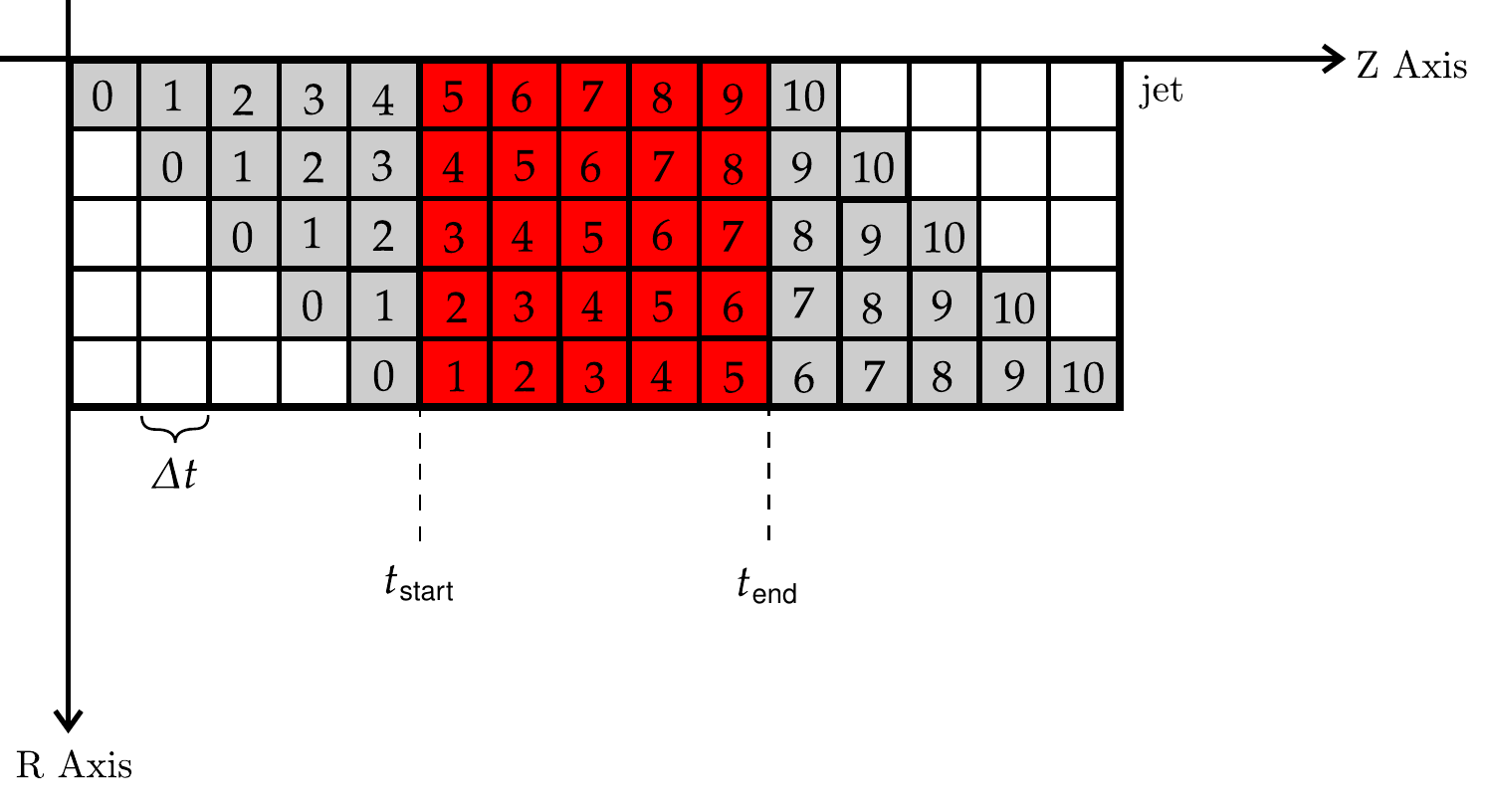}
    \caption{Schematic representation of our toy model. The red area represents the region where the shock is "on", gray when its "off". The number on each cells indicates the generation of electrons.}
    \label{fig: toy model LCE}
\end{figure}

We first tested the procedure on a simple toy model, similar to the one referred to by \cite{Chiaberge_1999}, which describes a flare event where a moving shock crosses a standing shock. The two-dimensional jet is modeled with square cells and the standing shock region corresponds to a small slice of a few cell columns (red region in Fig.~\ref{fig: toy model LCE}). The interaction is represented by the injection of electrons in all cells inside the shock region. As the the moving shock passes through the standing shock region, injection takes place first at the beginning of the region at time $t_{\rm start}$ and then also in consecutive columns. Injection stops everywhere, as the moving shock exits from the region at time $t_{\rm end}$. The post-processing code is directly applied to this model, assuming a viewing angle of $90^\circ{}$ in the observer frame. As electrons are injected in the cells moving through the shock
region, a flare appears in the synthetic synchrotron light curve. \\

In Fig.~\ref{fig: toy model LCE}, the numbers correspond to time steps in the observer frame. Light from cells with the same number reaches the observer at the same time. At time step 1, the observer starts detecting
emission from the onset of the shock interaction, but only from a position at the front edge of the jet and at the beginning of the shock region. For consecutive time steps, emission reaches the observer from
cells farther along the jet, as the moving shock crosses the standing shock region. At the same time, emission from cells closer to the jet
axis has propagated to the observer through the jet. This leads to the 
diagonal arrangement of cells with the same number. After the shock has
left the standing shock region, emission will continue during time steps 6, 7, 8, 9, as photons close to the jet axis arrive at the observer with an additional delay of $\Delta t_{\rm obs} = c \cdot l$ for each additional time step, where $l$ is the width of one cell. \\

To correctly account for this effect, before treating the ray tracing along each vertical column, the jet is split into layers, corresponding to the rows of cells in the toy model example. Each consecutive layer is shifted by one additional time step, such that, for example, photons emitted from the second layer propagate into the front layer, as it appears one time step later.

\begin{figure}[h]
    \centering
    \includegraphics[width = \columnwidth]{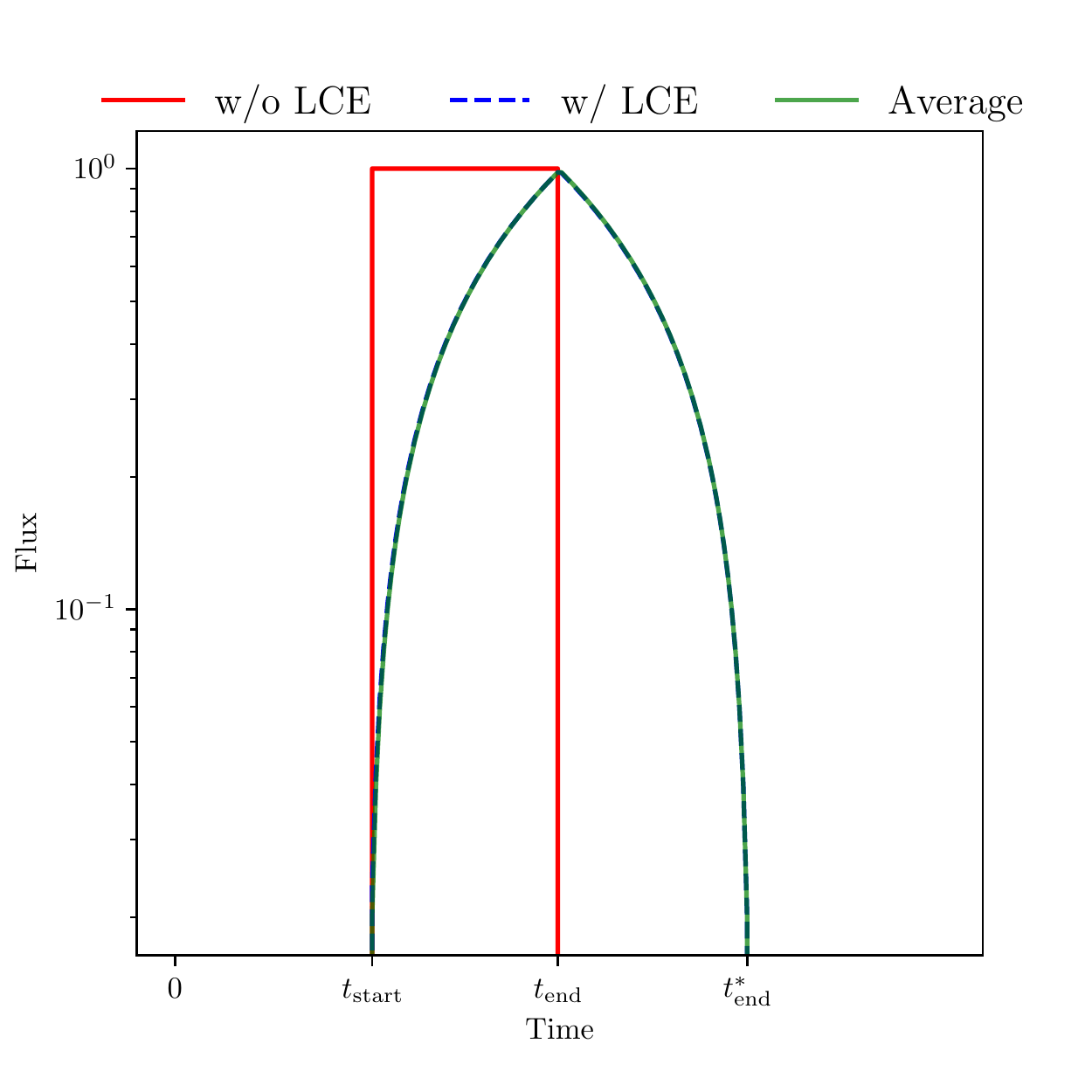}
    \caption{Light curves ($\nu = 10^{18}~\text{Hz}$) with and without LCE obtained for an observed situated in the co-mobile frame at $\theta_{\rm obs} = 90 \degree$.}
    \label{fig: LCE result}
\end{figure}

The resulting effect for the toy model is seen in Fig.~\ref{fig: LCE result}. Ignoring the LCE, the light curve is a simple Heaviside function with a width corresponding to the size of the injection region (red curve). Turning on the LCE, the number of ``activated'' cells for a given time step follows a moving average with a maximum at time step 5, when injection occurs over the whole stationary shock region. The result with a full treatment of the LCE is shown as the dashed blue curve. The change in shape is in good agreement with the result from \cite{Chiaberge_1999}
at high energies (see Fig. 6 in their paper).

For this particular case of a simple geometric emission region and a
$90^\circ$ observation angle, one can reproduce the LCE by applying a
simple moving average on the light curve, as is shown with the green 
curve, requiring far less computing time. The temporal width of this moving average must be equal to the light crossing time of one layer. 

\subsection{Application to full simulations for various $\theta_{\rm obs}$}

We first apply the ``layer method'' to full jet simulations for a viewing angle of $\theta_{\rm obs} = 90 \degree$ in the observer frame. The simulation time step needs to be sufficiently small, compared to the light crossing time of the jet radius, if one wants to resolve the LCE. Here we present results with $\Delta t'_{\rm obs} = 0.05~R_{\rm jet} / c$ (while for the general simulation we are using $1~R_{\rm jet} / c$). Due to the large amount of computing time, we have restricted the ray tracing treatment to emission coming from a restrained area of the jet centered on an X-ray flare event. 

At this angle, the LCE leads only to a small distortion and smoothing of the overall shape of the light curve, without a significant impact on the
observable structure.

When comparing the result with an application of a simple moving average (cf. \ref{subsection: LCE toy}), a good agreement is seen. The residuals between both methods show an average relative error of $\sim 1~\%$

\begin{figure}
    \centering
    \includegraphics[width = \columnwidth]{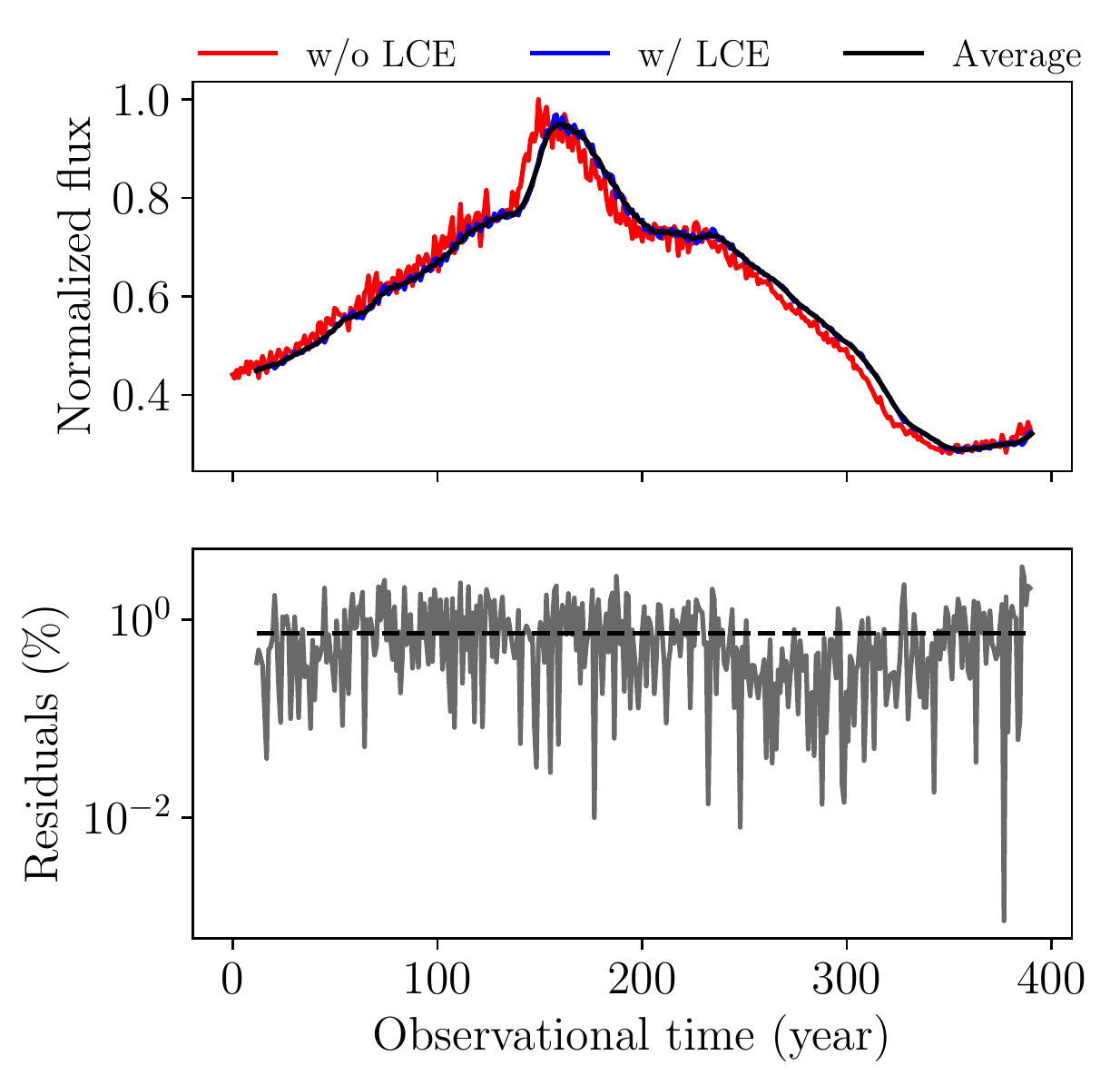}
    \caption{Light curves with (blue) and without (red) light crossing effect (LCE) for $10^{18}$ Hz at $\theta_{\rm obs} = 90 \degree$. The dark line represents the ``moving average method''. The lower panel shows the residuals (gray line, w/ LCE minus moving average over w/ LCE) between both methods.}
    \label{fig: with LTD 90}
\end{figure}

\begin{figure}
    \centering
    \includegraphics[width = \columnwidth]{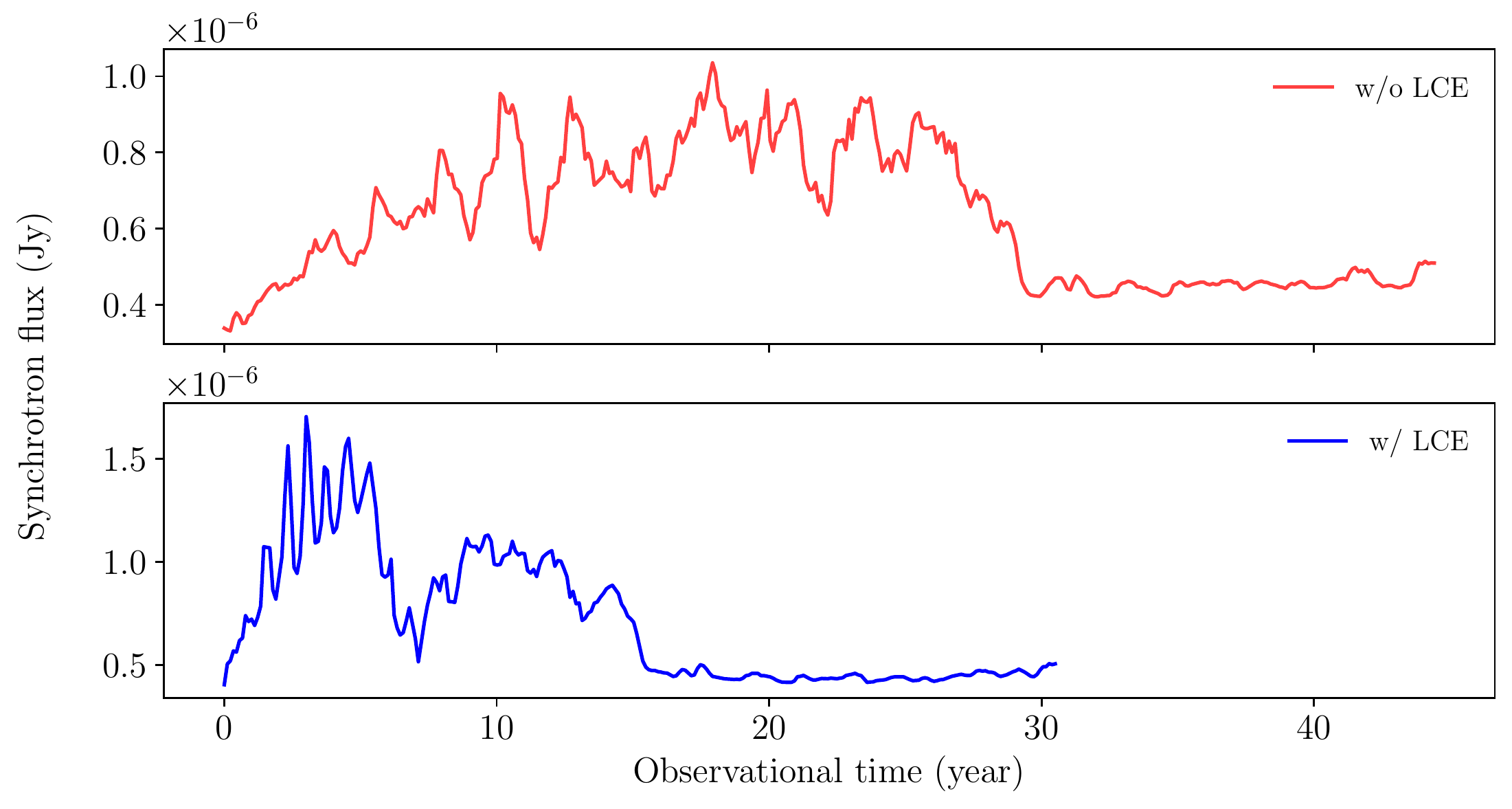}
    \caption{Light curve without and with LCE for $10^{18}$ Hz at $\theta_{\rm obs} = 20 \degree$. The flux is evaluated at the distance of the radio galaxy 3C\,111.}
    \label{fig: with LTD 20}
\end{figure}

We also present a result with a much smaller observation angle of $\theta_{\rm obs} = 20 \degree$, using the nominal simulation time step of $R_{\rm jet} / c$. We here compute the light curve by integrating the flux over the entire simulated jet, focusing again on the X-ray band. At this angle, the LCE has a significant impact on the observed light curve. We recover a temporal compression of the light curve, with flare activities concentrated in a period of $\sim 5~\text{yr}$. This shortened time scale is also reflected in Fig.~\ref{fig: nodes distance vs time 10}. Due to the apparent superluminal motion, the flare variability decreases to reach $150 - 200~\text{days}$. Additionally, the shape of the flares changes, but dedicated, computationally intensive simulations with smaller time steps are needed for a more detailed study, which is beyond the scope of this work. 

\section{Two further sources with potential trailing shocks.}
\label{sec: trailing candidates}

In this section, we present two further radio galaxies showing signs of the presence of trailing components in the core separation vs time plot, in the form of ``fork events'' connected to stationary knots. We use the MOJAVE database maintained by the MOJAVE team \citep{Lister_2018, Lister_2021}.

\begin{figure}[hbtp]
    \centering
    \includegraphics[width = \columnwidth]{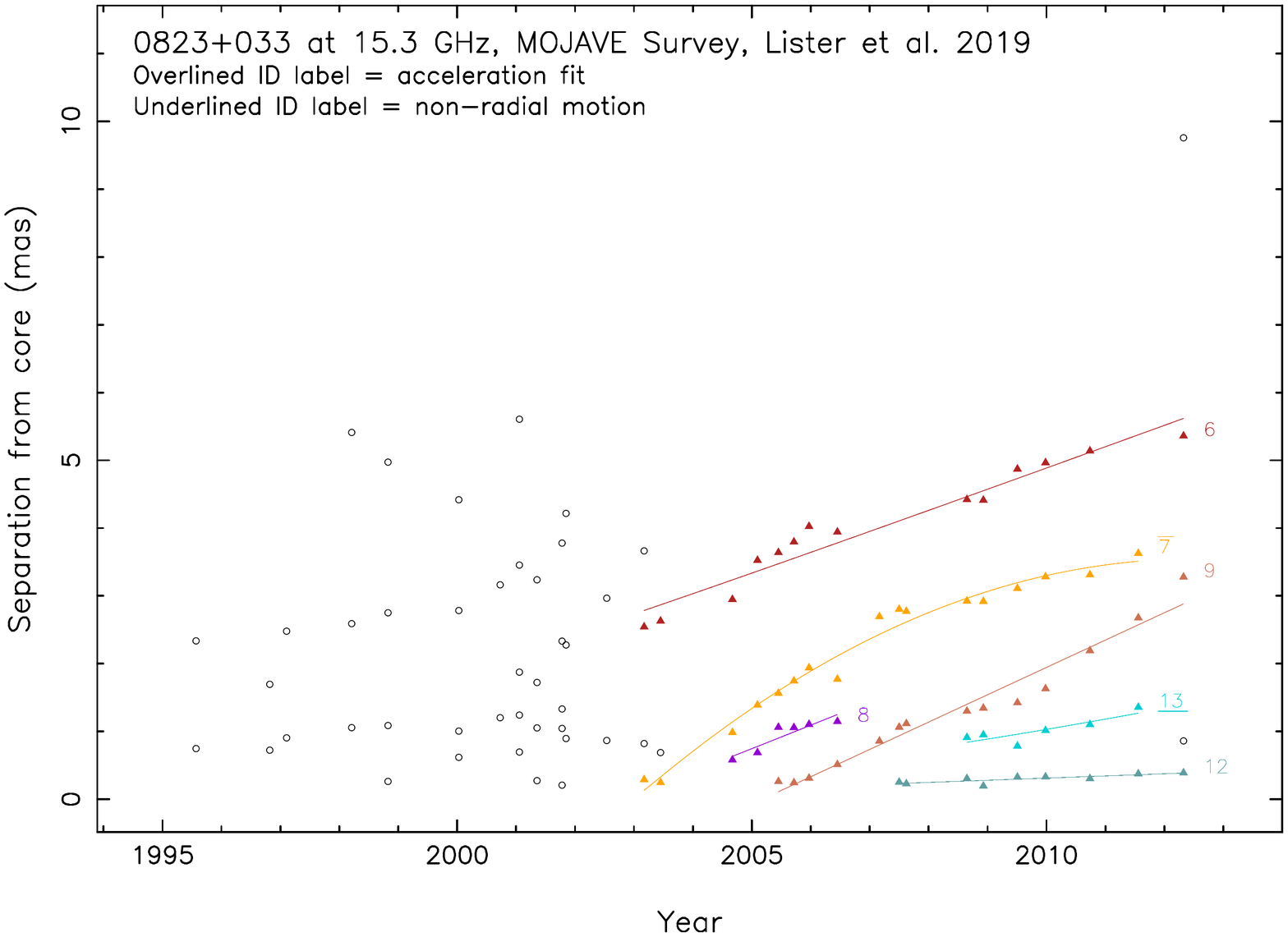}
    \caption{Core separation vs time for the source PKS 0823+033. Components 8 and 13 are candidates for relaxation shocks. A stationary shock is labeled as 12.}
    \label{fig: PKS 1546+27}
\end{figure}

\begin{figure}
    \centering
    \includegraphics[width = \columnwidth]{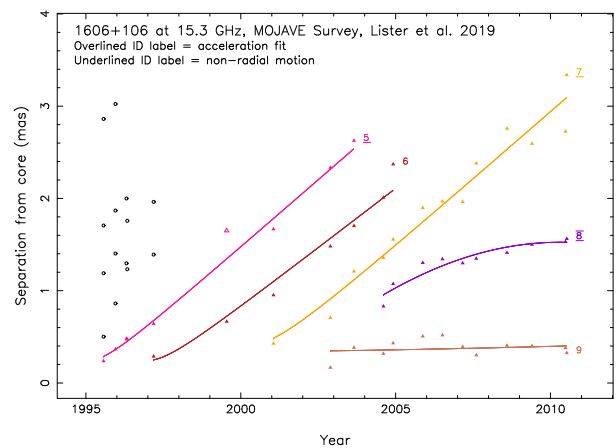}
    \caption{Core separation vs time for the source 4C +10.45. The component 8 may be a potential relaxation shock with component 9 as a stationary shock.}
    \label{fig: 4C +10}
\end{figure}

\end{document}